%% file: main.tex
\documentclass[11pt,twocolumn]{article}
\usepackage{graphicx}
\usepackage[left=1.80cm, right=1.80cm, top=2.00cm, bottom=2.00cm]{geometry}
\usepackage{amsmath,bm}
\usepackage[colorlinks]{hyperref}
\usepackage[natbib=true, sorting=none, backend=biber, bibencoding=utf8, style=ieee]{biblatex}
\usepackage{eurosym}
\usepackage[dvipsnames]{xcolor}
\usepackage{subcaption}
\usepackage{booktabs} 
\usepackage{enumitem} 
\usepackage{accents}
\usepackage{authblk}
\usepackage{tikz}
\usepackage{dblfloatfix}
\usepackage[printonlyused]{acronym}
\usepackage{cleveref}

\graphicspath{{figures/}}

\usetikzlibrary{arrows}
\tikzset{
    vertex/.style={circle,draw,minimum size=1.5em},
    edge/.style={->,> = latex',very thick}
}

\crefname{relation}{Rel.}{Rels.}
\creflabelformat{relation}{(#2#1#3)}
\crefname{constraint}{Constr.}{Constrs.}
\creflabelformat{constraint}{(#2#1#3)}
\crefname{subequations}{Eqs.}{Eqs.}
\creflabelformat{subequations}{(#2#1#3)}

\newcommand{\ie}{i.e.}
\newcommand{\eg}{\textit{e.g.} }
\newcommand{\ubar}[1]{\underaccent{\bar}{#1}}

\newcommand{\resultsin}[1]{\hspace{6pt} \leftrightarrow  \hspace{6pt} #1}
\newcommand{\Forall}[1]{\hspace{10pt} \forall \,\, #1 }
\newcommand{\pdv}[2]{\frac{\partial #1}{\partial #2}}
\newcommand{\ra}{\rightarrow}

\newcommand{\capacity}{S_{i}}

\newcommand{\capacityupper}{\bar{S}}
\newcommand{\capacitylower}{\ubar{S}}
\newcommand{\muuppernom}{\bar{\mu}^\text{nom}_{i}}
\newcommand{\mulowernom}{\ubar{\mu}^\text{nom}_{i}}

\newcommand{\generation}{g_{s,t}}
\newcommand{\generationpotential}{\bar{g}_{s,t}}

\newcommand{\nodalgeneration}[1][n]{g_{#1,t}}
\newcommand{\capacitygeneration}{G_{s}}

\newcommand{\operationalpricegeneration}{o_{s}}
\newcommand{\capitalpricegeneration}{c_{s}}
\newcommand{\mulowergeneration}{\ubar{\mu}_{s,t}}
\newcommand{\muuppergeneration}{\bar{\mu}_{s,t}}
\newcommand{\muuppergenerationnom}{\bar{\mu}^\text{nom}_{s}}

\newcommand{\flow}{f_{\ell,t}}
\newcommand{\capacityflow}{F_{\ell}}

\newcommand{\capitalpriceflow}{c_{\ell}}
\newcommand{\mulowerflow}{\ubar{\mu}_{\ell,t}}
\newcommand{\muupperflow}{\bar{\mu}_{\ell,t}}

\newcommand{\storage}{g_{r,t}}
\newcommand{\storagedispatch}{\storage^\text{dis}}
\newcommand{\storagecharge}{\storage^\text{sto}}
\newcommand{\storagesoc}{\storage^\text{ene}}
\newcommand{\storageprevioussoc}{g_{r,t-1}^\text{ene}}

\newcommand{\efficiency}{\eta_{r}}
\newcommand{\efficiencydispatch}{\efficiency^\text{dis}}
\newcommand{\efficiencycharge}{\efficiency^\text{sto}}
\newcommand{\efficiencysoc}{\efficiency^\text{ene}}

\newcommand{\operationalpricestorage}{o_r}
\newcommand{\capitalpricestorage}{c_r}
\newcommand{\capacitystorage}{G_r}

\newcommand{\mulowerstoragedispatch}{\ubar{\mu}_{r,t}^\text{dis}}
\newcommand{\muupperstoragedispatch}{\bar{\mu}_{r,t}^\text{dis}}

\newcommand{\muupperstoragecharge}{\bar{\mu}_{r,t}^\text{sto}}

\newcommand{\muupperstoragesoc}{\bar{\mu}_{r,t}^\text{ene}}

\newcommand{\mustateofcharge}{\lambda^\text{ene}_{r,t}}

\newcommand{\lagrangian}{\mathcal{L}}
\newcommand{\lmp}[1][n]{\lambda_{#1,t}}
\newcommand{\lmpdiff}[1][\ell]{\lmp[#1]^\text{diff}}
\newcommand{\lmpkvl}[1][\ell]{\lmp[#1]^\text{KVL}}
\newcommand{\averagelmp}[1][n]{\bar{\lambda}_{#1}}
\newcommand{\demand}[1][n]{d_{#1,t}}

\newcommand{\netconsumption}[1][n]{p^{-}_{#1,t}}
\newcommand{\netproduction}[1][n]{p^{+}_{#1,t}}

\newcommand{\incidence}[1][n]{K_{#1,\ell}}
\newcommand{\incidencegenerator}[1][n]{K_{#1,s}}
\newcommand{\incidencestorage}[1][n]{K_{#1,r}}

\newcommand{\ptdf}[1][n]{H_{\ell,#1}}
\newcommand{\cycle}{C_{\ell,c}}
\newcommand{\reactance}{x_\ell}
\newcommand{\cycleprice}{\lambda_{c,t}}

\newcommand{\emission}{e_{s}}
\newcommand{\emissionprice}{\mu_{\text{CO2}}}
\newcommand{\megawatthour}{MWh$_\text{el}$}

\newcommand{\cost}{\mathcal{C}}
\newcommand{\payment}[1][n]{\cost_{#1,t}}

\newcommand{\opex}{\mathcal{O}}
\newcommand{\opexgeneration}{\mathcal{O}^G}

\newcommand{\opexstorage}{\mathcal{O}^E}
\newcommand{\capex}{\mathcal{I}}
\newcommand{\capexgeneration}{\mathcal{I}^G}
\newcommand{\capexflow}{\mathcal{I}^F} 
\newcommand{\capexstorage}{\mathcal{I}^E}
\newcommand{\emissioncost}{\mathcal{E}}
\newcommand{\scarcitycost}{\mathcal{S}}
\newcommand{\subsidycost}{\mathcal{U}}
\newcommand{\remainingcost}{\mathcal{R}}

\newcommand{\allocategeneration}[1][s, n]{A_{#1,t}}
\newcommand{\allocatestoragedispatch}[1][r, n]{A_{#1,t}}
\newcommand{\allocatepeer}[1][m \rightarrow n]{A_{#1,t}}
\newcommand{\allocateflow}[1][n]{A_{\ell,#1,t}}

\newcommand{\allocatecost}[1][n \rightarrow i]{\cost_{#1,t}}

\newcommand{\allocategeneratorcost}[1][n \rightarrow s]{\cost_{#1, t}}
\newcommand{\allocatelinecost}[1][n \rightarrow \ell]{\cost_{#1, t}}

\newcommand{\allocatecapex}[1][n \rightarrow s,t]{\mathcal{I}_{#1}}

\newcommand{\allocateopex}[1][n \rightarrow s]{\opex_{#1,t}}
\newcommand{\allocateemissioncost}[1][n \rightarrow s]{\emissioncost_{#1,t}}
\newcommand{\allocatescarcitycost}[1][n \rightarrow s]{\scarcitycost_{#1,t}}

\begin{document}

\title{Price Tracing: Linking Nodal Prices in Optimized Power Systems}
\author[1,2]{Fabian Hofmann}
\author[1]{Markus Schlott}

\affil[1]{Frankfurt Institute for Advanced Studies (FIAS), 60438 Frankfurt, Germany}
\affil[2]{Technical University Berlin, Department of Energy Systems, 10623 Berlin, Germany}

\date{}

\twocolumn[
    \begin{@twocolumnfalse}

        \maketitle

        \begin{abstract}
            Optimizing the total cost of power systems is a common tool for network operation and planning. Besides valuable information about how to run and possibly expand a power system, the optimization provides an optimal Locational Marginal Price per node and time step. This price can be seen as the price of electricity paid by consumers and purchased by suppliers, while maximizing social welfare. Naturally, it is a direct result of the optimization problem, and therefore does not give any information about its internal composition. This paper shows that by applying Flow Tracing, an algorithm for tracking flows in complex networks, it is possible to interlink Locational Marginal Prices in a coherent mathematical way. This does not only lead to important insights into the price structure, but also provides an intuitive decomposition and allocation of all system costs. Then individual consumers ``see'' how much they have to pay to individual generators and transmission lines in the power system. This method, introduced as Price Tracing, outperforms similar approaches provided by the literature, since the resulting cost allocations are transparent, plausible, and consistent with the Locational Marginal Prices from the optimization. The Price Tracing method is applied and discussed on behalf of a power system model of Germany with a high share of renewable power. The presented analysis and its implications can help in finding a more efficient market design to promote renewable power supply.
        \end{abstract}

    \end{@twocolumnfalse}
]

\subsection*{Highlights}
\begin{itemize}
    \item Flow Tracing is used to interlink optimal Locational Marginal Prices.
    \item This quantifies the payment of individual consumers to individual generators and transmission lines.
    \item Four use-cases are proposed and demonstrated on behalf of a future model of the German power system.
\end{itemize}



\section{Introduction}

Today's energy systems are undergoing profound and enduring change. The transition from controllable to variable, weather-dependent power generation and the constant improvement and innovation of technology require rigorous system planning and international cooperation. The core of the challenge manifests itself in the total cost of the system. First, these should be as low as possible, while meeting environmental and techno-economic standards. Second, they must be distributed fairly and transparently among all actors in the power system. It is essential to identify the cost drivers and account for them appropriately.
In this regard, power system models are a valuable and widely used tool \cite{pfenninger_energy_2014,bazmi_sustainable_2011,pereira_generation_2017,brown_sectoral_2019}. Many studies looking at different countries and regions around the world show how to expand renewable energy penetration at minimal cost. However, they are largely silent on how and for what reasons these costs are passed on to consumers.

This paper addresses this gap. We introduce a new method called Price Tracing that links electricity prices from different nodes. The method builds on Bialek's \ac{AP} \cite{bialek_tracing_1996}, also known as flow tracing, and weights the resulting power flow allocations with the \ac{LMP}. This provides the basis for a transparent allocation of cost contributions to consumer electricity prices, i.e. it answers not only the question of who supplies electricity to whom, but also how much consumers must pay to particular generators or transmission lines.

In the literature, the concept of flow-based cost allocation has been discussed and applied in a number of works \cite{galiana_transmission_2003,shahidehpour_market_2002,meng_investigation_2007,schafer_allocation_2017,nikoukar_transmission_2012,arabali_pricing_2012,wu_locational_2005}. Shahidehpour et al. provide deep insights into the allocation of congestion costs and transmission investments to market participants using different allocation techniques \cite{shahidehpour_market_2002}. In particular, they suggest that generation shift factors, i.e. the marginal contribution of generators to a flow on a line, allow the \ac{LMP} to be represented as a superposition of (a) the \ac{LMP} on the reference bus, (b) the congestion price, and (c) a price for losses. The approach in \cite{meng_investigation_2007} extends this relationship for contributions based on the \ac{AP} scheme, which provides an accurate, albeit imprecise, estimate of the optimal \ac{LMP}. A similar approach based on \ac{AP} is used in \cite{tranbergFlowbasedNodalCost2018}, which allocates electricity prices from a non-optimal electricity dispatch.

The presented approach combines the advantages of the above studies. It ensures that cost allocations are plausible and spatially constrained, while payments are fully aligned with the \ac{LMP}-based nodal pricing system. It also facilitates transparency and cost-benefit analysis in network planning, such as the ten-year network development plan \cite{entso-e_completing_2020} or the German network development plan \cite{bundesnetzagentur_netzentwicklungsplan_2020}. In addition, it provides a starting point for a usage-based allocation of transmission costs. While the presented method uses the \ac{LOPF} to analyze large network models, it is discussed how it can be extended to nonlinear power flows.

First, we formulate the price-tracing method (\cref{sec:price_tracing}) with the underlying assumptions and a numerical example (\cref{sec:numerical_example}). In \cref{sec:application_case}, possible use cases are presented using an optimized German power system with a high share of renewable resources. \Cref{sec:limitations} shows methodological limitations and \cref{sec:conclusions} draws final conclusions.

\section{Price Tracing}
\label{sec:price_tracing}

\subsection{Mathematical Model}

Consider an electricity network model with N nodes, L lines and T time steps. Using the linearized power flow approximation, the power flow $\flow$ on a passive line $\ell$ at time $t$ relates to the generation $\nodalgeneration$ and demand $\demand$ at node $n$ and time $t$ according to
\begin{align}
    \flow = \sum_n \ptdf \left(\nodalgeneration - \demand \right) \Forall{t}
\end{align}
where $\ptdf$ are the \ac{PTDF}. These translate the nodal injection on the right-hand side to the network flow in compliance with the linearized Kirchhoff Circuit Laws.

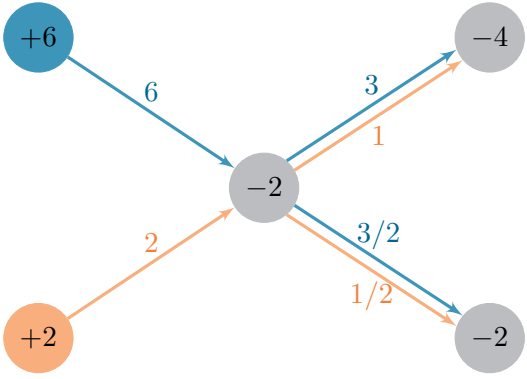
\begin{figure}
    \centering
    \begin{tikzpicture}

        \node[vertex, draw=Orange!60, fill=Orange!60] (1) at (0,0) {$+2$};
        \node[vertex, draw=MidnightBlue!60, fill=MidnightBlue!60] (2) at (0,4) {$+6$};
        \node[vertex, draw=Gray!60, fill=Gray!60] (3) at (3,2) {$-2$};
        \node[vertex, draw=Gray!60, fill=Gray!60] (4) at (6,0) {$-2$};
        \node[vertex, draw=Gray!60, fill=Gray!60] (5) at (6,4) {$-4$};

        \draw[edge, draw=Orange!60] (1) -- (3) node[midway, above] {\color{Orange}$2$};
        \draw[edge, draw=MidnightBlue!60] (2) -- (3) node[midway, above] {\color{MidnightBlue}$6$};
        \draw[edge, draw=MidnightBlue!60] (3.330) -- (4.140) node[midway, above] {\color{MidnightBlue}$3/2$};
        \draw[edge, draw=Orange!60] (3.310) -- (4.160) node[midway, below] {\color{Orange}$1/2$};
        \draw[edge, draw=MidnightBlue!60] (3.050) -- (5.200) node[midway, above] {\color{MidnightBlue}$3$};
        \draw[edge, draw=Orange!60] (3.030) -- (5.220) node[midway, below] {\color{Orange}$1$};
    \end{tikzpicture}
    \caption{Schematic illustration of the \ac{AP} method. Figuratively, each power source is given one color. As soon as different power flows mix at a node, the mix of ``colors'' in the outgoing flows is in proportion to the ingoing colors. Therefore, all consumed power can be related to a distinct origin.}
    \label{fig:ap-scheme}
\end{figure}

In a nodal pricing scheme, the system cost occasioned by the electrical demand $\demand$ is proportional to the corresponding \ac{LMP} $\lmp$,
\begin{align}
    \text{Demand Cost}: \lmp\, \demand .
    \label{eq:demand_cost}
\end{align}
On the other hand, system assets generate revenue from their operation. Therefore, the revenue of the dispatch $\nodalgeneration[m]$ at bus $m$ is given by
\begin{align}
    \text{Dispatch Revenue}: \lmp[m] \, \nodalgeneration[m]
    \label{eq:dispatch_revenue}
\end{align}
and the congestion revenue of line $\ell$ at time $t$ by
\begin{align}
    \text{Congestion Revenue}: \lmp[\ell]\, \flow .
    \label{eq:congestion_revenue}
\end{align}
where, in the absence of network cycles, the revenue per transported MWh $\lmp[\ell]$ is the price difference $\lmpdiff$ between the end node and beginning node of line $\ell$. In the case of network cycles, the price of the \ac{KVL} $\lmpkvl$ may be added to $\lmp[\ell]$ to match congestion revenues to expenditures in a long-term equilibrium, see \cref{sec:problem_formulation,sec:zero_profit_flow} for details. The \ac{LMP} as well as the prices for the \ac{KVL} are given by the dual values, often called shadow prices, of the corresponding constraints in the underlying cost optimization (see \cref{sec:optimality_conditions} for details).
In economic equilibrium, prices naturally arise, which in turn leads to the following equivalency
\begin{align}
    \underbrace{\sum_{n} \lmp\, \demand}_{\text{Total electricity cost at $t$}}  = \underbrace{\sum_{m} \lmp[m]\, \nodalgeneration[m] + \sum_{\ell} \lmp[\ell] \flow}_{\text{Total revenue at $t$}}\, \Forall{t} .
    \label{eq:total-demand-cost}
\end{align}
While this equality relates the totals of costs and revenues, the relationships between individual contributions on the left-hand side and individual contributions on the right-hand side remain undefined. The following shows that by allocating flows in the network, it is possible to relate the power consumption at single nodes to the revenues of single generators and transmission lines:

Dispatch and flow can be considered as a superposition of individual contributions of nodes and assets. The literature provides various methods, called \ac{FA} method, to artificially quantify these. Each method follows a particular set of assumptions that lead to \ac{P2P} allocations $A_{m \rightarrow n}$ which measure the power generated at node $m$ and consumed at node $n$. Flow Tracing, also known as \ac{AP} \cite{bialek_tracing_1996}, is a flow allocation method that traces the power injection at bus $m$ through the network up to its sink $n$ using the principle of proportional sharing. The method is illustrated in \cref{fig:ap-scheme} and mathematically documented in \cref{sec:net_ap}.
It states that when flows from different sources meet at the same bus, their share determines the mixture of all flows originating from that bus, including nodal withdrawal. The advantage of the \ac{AP} method is the spatial containment of \ac{P2P} allocations $A_{m \rightarrow n}$ \cite{hofmann_techno-economic_2020}. By definition, the sum of all receivers gives the nodal generation at $m$, thus
\begin{align}
    \nodalgeneration[m] = \sum_n \allocatepeer \Forall{m.t}
    \label{eq:nodalgeneration-breakdown}
\end{align}
As explained later in the paper, only the \ac{P2P} mappings of the \ac{AP} method are used. Thus, regardless of the \ac{AP} allocation, $\allocateflow$ denotes the contribution of demand $\demand$ to the flow on line $\ell$ such that the sum of all flow contributions equals flow on line $\ell$,
\begin{align}
    \flow = \sum_n \allocateflow \Forall{\ell,t}.
    \label{eq:flow-breakdown}
\end{align}
Inserting \cref{eq:nodalgeneration-breakdown,eq:flow-breakdown} into \cref{eq:total-demand-cost} and imposing that the equality holds for all summands referring to $n$ separately, leads to
\begin{align}
    \lmp\, \demand & = \sum_m \lmp[m] \allocatepeer + \sum_\ell \lmp[\ell] \allocateflow \Forall{n,t}.
    \label{eq:demand-cost-allocation}
\end{align}
This equation maps the cost of nodal demand on the left-hand side to the contributions of dispatch and congestion revenues on the right-hand side. It thus indicates what consumers at $n$ have to pay to generators at $m$ and the transmission line $\ell$. The equation introduces N new equalities for each time step for which the necessary degree of freedom comes from the yet undefined flow contribution $\allocateflow$. \Cref{sec:proof_equivalence} proves that
\begin{align}
    \allocateflow = \sum_m \ptdf[m] (\allocatepeer - \delta_{nm} \demand)
    \label{eq:allocateflow}
\end{align}
solves \cref{eq:demand-cost-allocation} where $\delta_{mn}$ is 1 for $m=n$ and zero otherwise. Since \cref{eq:allocateflow} is a function of the \ac{P2P}-assignment $\allocatepeer$ only, the cost of electricity in \cref{eq:demand-cost-allocation} is also only a function of the \ac{P2P} allocation $\allocatepeer$ and the \ac{LMP} $\lmp$.
Note that it is not possible to take the non-modified flow allocations given by the \ac{AP} as these are not a solution to \cref{eq:demand-cost-allocation}. Instead it is necessary to reassign them through \cref{eq:allocateflow}\footnote{This can be explained by considering that the \ac{AP} flow allocations do not respect the \ac{KVL}, which however have a direct effect on the \ac{LMP}. For the same reason, the cost estimation in \cite{meng_investigation_2007} are also inaccurate.}.

Since generators are often grouped together at network nodes, the \ac{P2P}-assignments $\allocatepeer$ partition into the contributions of generators located at $m$. Following the formulation in \cite{schafer_tracing_2020}, let $\allocategeneration = \allocatepeer \, r_{s,m,t}$ be the power allocation from generator $s$ to demand at $n$ where $r_{s,m,t}$ is the share that generation $\generation$ of generator $s$ at time $t$ contributes to the generation $\nodalgeneration[m]$ at node $m$ and $t$. It fulfills
\begin{align}
    \generation = \sum_n \allocategeneration .
    \label{eq:generation-breakdown}
\end{align}
Further, let $\lmp[s]$ of generator $s$ be set to the \ac{LMP} $\lmp$, where $s$ is located at bus $n$. For consistency with \cref{eq:dispatch_revenue}, the dispatch revenue of generator $s$ is given by $\lmp[s] \, \generation$.\

The \textit{Price Tracing} method for a network with a linearized power flow and a nodal pricing scheme builds on the above equations and defines
\begin{subequations}
    \begin{align}
        \allocategeneratorcost = \lmp[s] \allocategeneration
    \end{align}
    as the payment from consumers at $n$ to generator $s$ as well as
    \begin{align}
        \allocatelinecost = \lmp[\ell] \allocateflow
    \end{align}
\end{subequations}
as the payment to transmission line $\ell$. For consistency, let $\payment = \lmp \demand$ denote the nodal payment of all consumers at node $n$ which according to \cref{eq:demand-cost-allocation} is $\payment = \sum_s \allocategeneratorcost + \sum_\ell \allocatelinecost$.

It should be noted that storage units such as batteries can easily be included in the Price Tracing method: When they discharge electricity, they are treated as generators, and when they charge electricity, they are treated as consumers. For the latter, the $\allocategeneration$ and $\allocateflow$ allocations must be split into allocations to consumers and charging storage units at $n$.


\subsection{Numerical Example}
\label{sec:numerical_example}

In the following, the Price Tracing method is illustrated by a numerical example, assuming a three-node network with two lines. Considering one time step, the dispatch of the generators is optimized and the costs are allocated. Finally, a third line is added to the system and the cost allocation is repeated in order to show the effects of the network cycle.

\subsubsection*{Network without Cycles}
\begin{figure}[h!]
    \includegraphics[width=\linewidth]{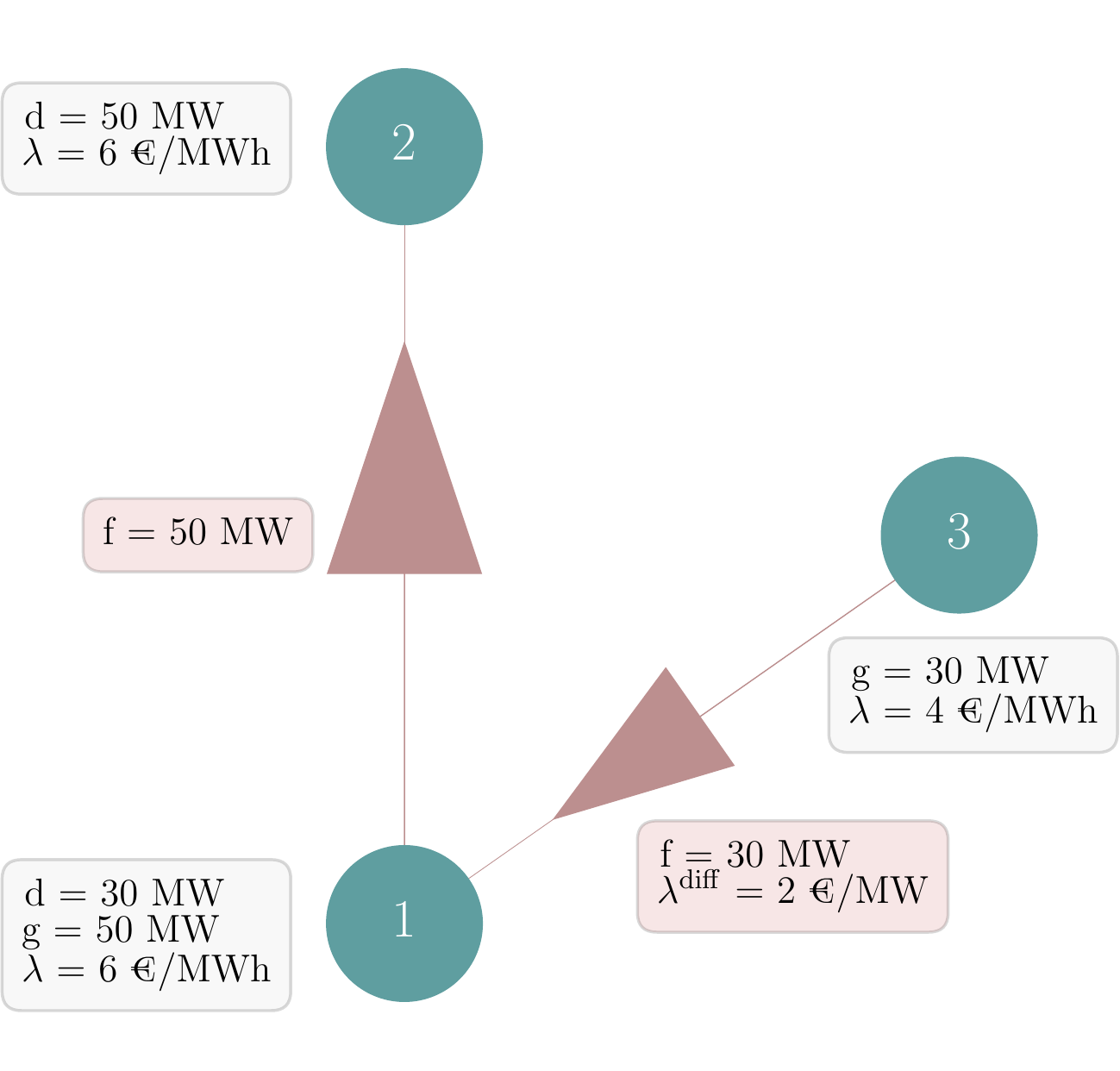}
    \caption{Example network with nodal pricing scheme. Dispatch revenue as well as demand cost per MWh are given by the \ac{LMP} $\lambda$. Congestion revenue per MWh is given by the price difference $\lambda^\text{diff}$ of the connecting nodes.}
    \label{fig:example-network-wo}
\end{figure}

\Cref{fig:example-network-wo} shows the optimized three-node network with corresponding numerical values.
Both, bus 1 and 2 have a fixed demand of 30~MW and 50~MW, respectively. Bus~1 has a generator with an operational price of 6~\euro/MWh. Bus~3 has a cheaper generator with an operational price of 4~\euro/MWh. The maximum capacity of both generators is 50~MW. The transmission line from 1 to 2 has a capacity of 60~MW and line from 3 to 1 a capacity of 30 MW. Since the latter limits the use of the cheaper generator, there is a price difference of 2 \euro/MWh between bus 1 and 3. Due to the absence of network cycles, this price difference defines the congestion revenue of line 1--3. The flow on line 1--2 is not bound, hence there is no congestion revenue and bus 2 and 1 have the same electricity price of 6~\euro/MWh.

\begin{figure}[h]
    \includegraphics[width=\linewidth]{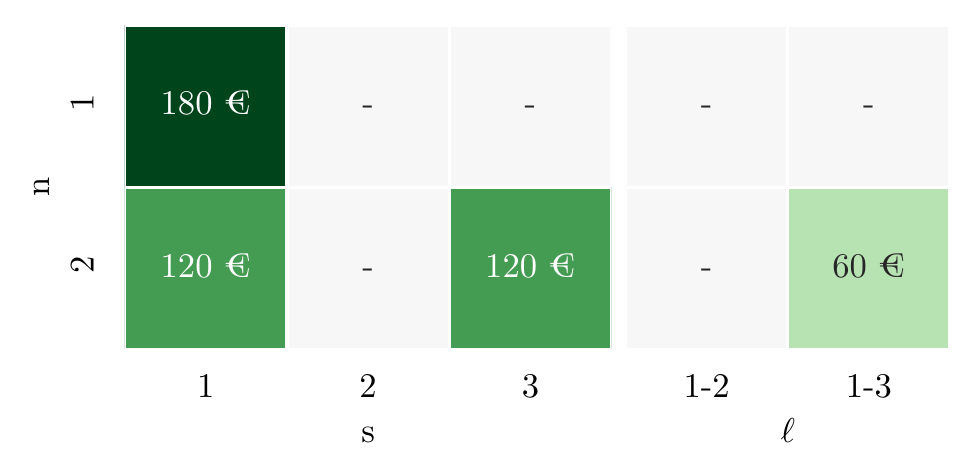}
    \caption{Allocation matrices $\mathcal{C}_{n\rightarrow s}$ (left) and $\mathcal{C}_{n\rightarrow \ell}$ (right) for the dispatch and congestion revenue for the network in \cref{fig:example-network-wo}.}
    \label{fig:example-network-payoff-wo}
\end{figure}

Since only net injections are allocated to the different buses, all demand at bus~1 is met by the local generator. Bus~2, on the other hand, imports electricity from bus~1 and bus~3. Using the Price Tracing method, the cost allocations $\mathcal{C}_{n\rightarrow s}$ and $\mathcal{C}_{n\rightarrow \ell}$ are calculated and shown in \cref{fig:example-network-payoff-wo}. Consumers at bus~1 only have to compensate the local generator with 180 \euro. Consumers at bus~2 pay for congestion revenues from line 1--3, all electricity generation at bus~3, and the remaining dispatch revenues from the generator at bus~1.

The consistency of the cost allocation can be easily checked. The sum of a column gives the payment to a generator or a line. This corresponds exactly to the dispatch and congestion revenue defined in \cref{eq:dispatch_revenue,eq:congestion_revenue}. In turn, the sum of a row gives the total amount consumers have to pay at a bus. This corresponds to the demand cost $\lambda_n \, d_n$, \eg the sum of payments to bus~2 is 300~\euro, equal to the price of 6~\euro/MWh times the 50~MWh consumption. \\

\subsubsection*{Network with Cycle}

\begin{figure}[h!]
    \includegraphics[width=\linewidth]{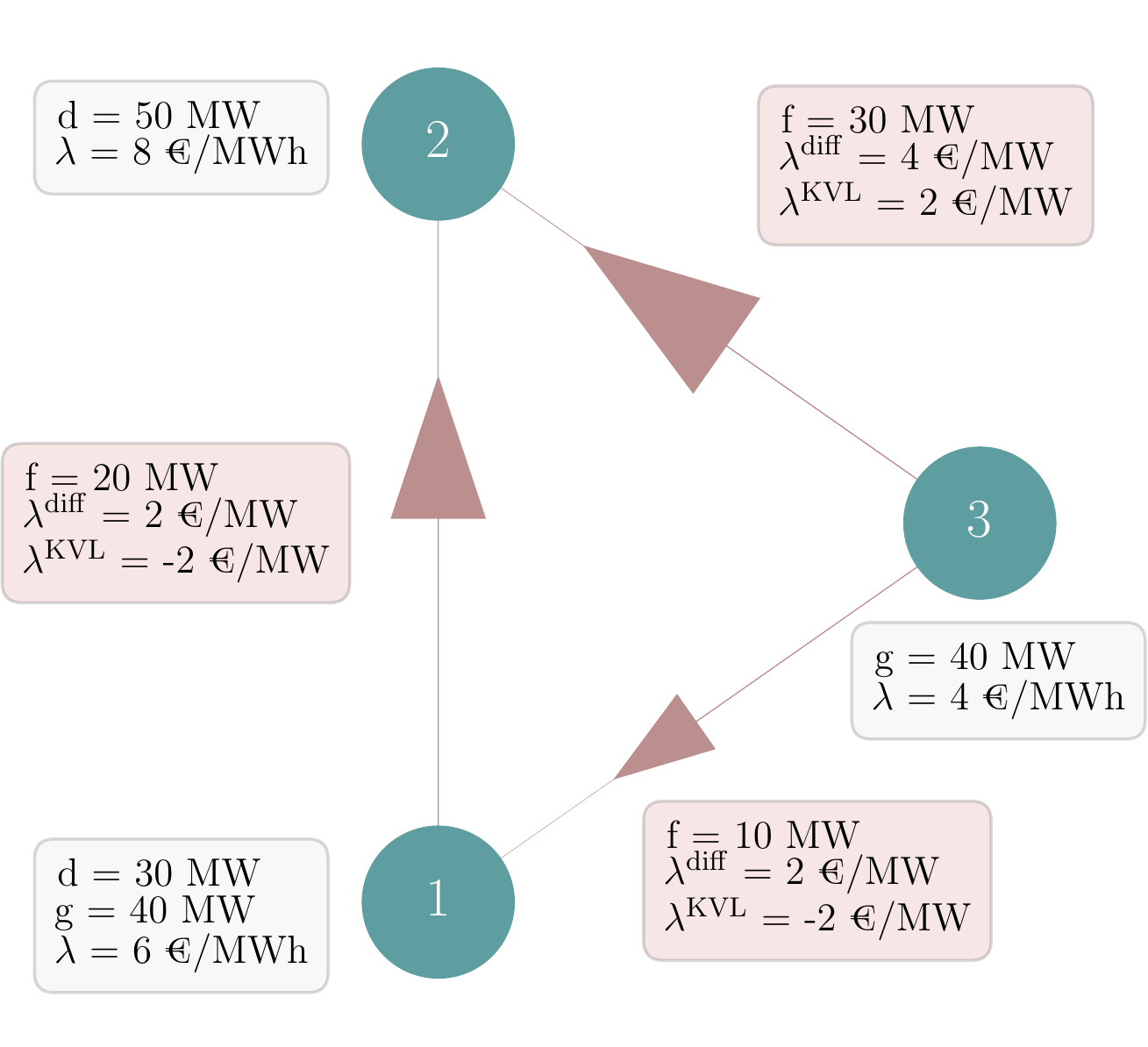}
    \caption{Example network with nodal pricing scheme and a network cycle. Now, due to the \ac{KVL} a new constraint adds to the optimization problem. Optionally, the congestion revenue per MWh may include shadow prices of the \ac{KVL} constraint $\lmpkvl$, thus $\lmp[\ell] = \lmpdiff + \lmpkvl$.}
    \label{fig:example-network}
\end{figure}

Now, we add a line from bus~3 to bus~2 with a capacity of 30~MW. This introduces \ac{KVL} constraints for the three lines that the sum of flows following the cycle is zero\footnote{For simplicity, we assume a uniform impedance in the example.}. As shown in \cref{fig:example-network} the new line's capacity is bound at 30~MW. The new cost optimum results in a higher price at bus~2 of 8~\euro/MWh despite the cheaper generator~2 producing 10~MW more than before. This result may be counterintuitive, but it becomes clear when considering that the low capacity of the new line, combined with the \ac{KVL}, restricts flows on line 1--3 and 1--2.

\begin{figure}[h!]
    \begin{subfigure}[c]{\linewidth}
        \includegraphics[width=\linewidth]{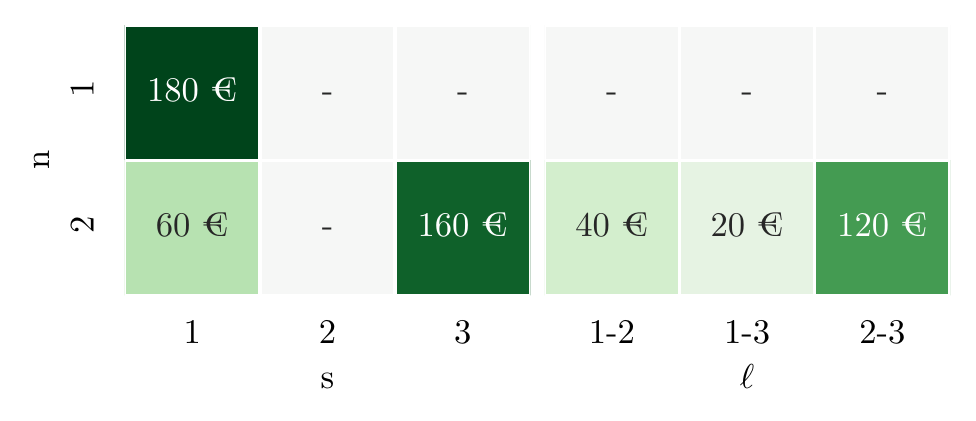}
        \subcaption{Without \ac{KVL} shadow prices}
        \label{fig:example-network-payoff}
    \end{subfigure}
    \begin{subfigure}[c]{\linewidth}
        \includegraphics[width=\linewidth]{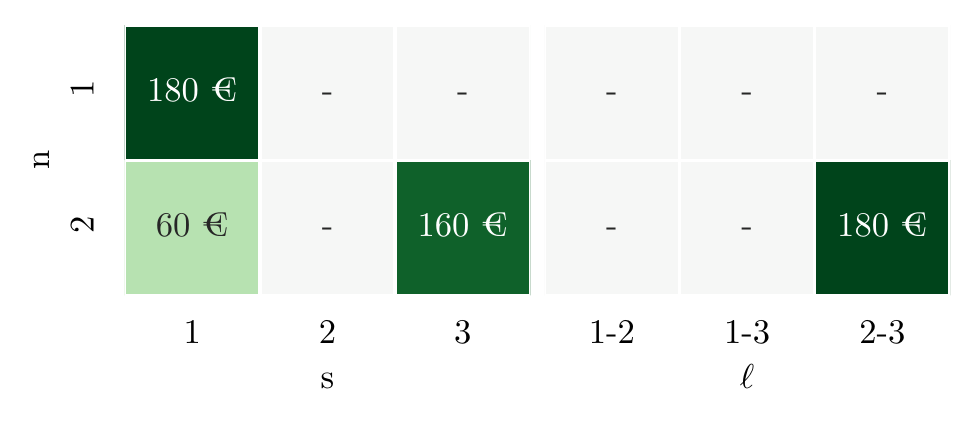}
        \subcaption{With \ac{KVL} shadow prices}
        \label{fig:example-network-payoff-kvl}
    \end{subfigure}
    \caption{Cost allocations $\allocategeneratorcost$ and $\allocatelinecost$ for the dispatch and congestion revenue in the network in \cref{fig:example-network}.}
    \label{fig:example-network-payoff-all}
\end{figure}

The prices $\lambda_\ell^\text{KVL}$ of line $\ell$ are returned as the shadow price of the \ac{KVL} constraints in the cost-optimization, see \cref{sec:problem_formulation} for details. The Price Tracing method optionally allows these to be included in the congestion revenue without losing consistency with the demand cost.

\Cref{fig:example-network-payoff} shows the cost allocation without considering the \ac{KVL} shadow prices. Here, the congestion revenue is proportional to the price difference of the connected buses only. The new line 2--3 has the highest revenue at 120~\euro/MWh. However, \cref{fig:example-network-payoff-kvl} shows the cost allocation taking into account the \ac{KVL} shadow prices. Here, the congestion revenues are shifted towards 1--2 and 1--3. Note that this shift accounts for the fact that it is the new line 2--3 that limits the flow on line 1--2 and 1--3.

Again, both cost allocation schemes are consistent with the total demand cost, dispatch and congestion revenues. In the following, the \ac{KVL} shadow prices in the Price Tracing method are considered since then, the total congestion revenue of a line equals the sum of all associated line costs in a short and long-term equilibrium, see \cref{sec:zero_profit_flow}.

\begin{figure*}[h!]
    \centering
    \includegraphics[width=\linewidth]{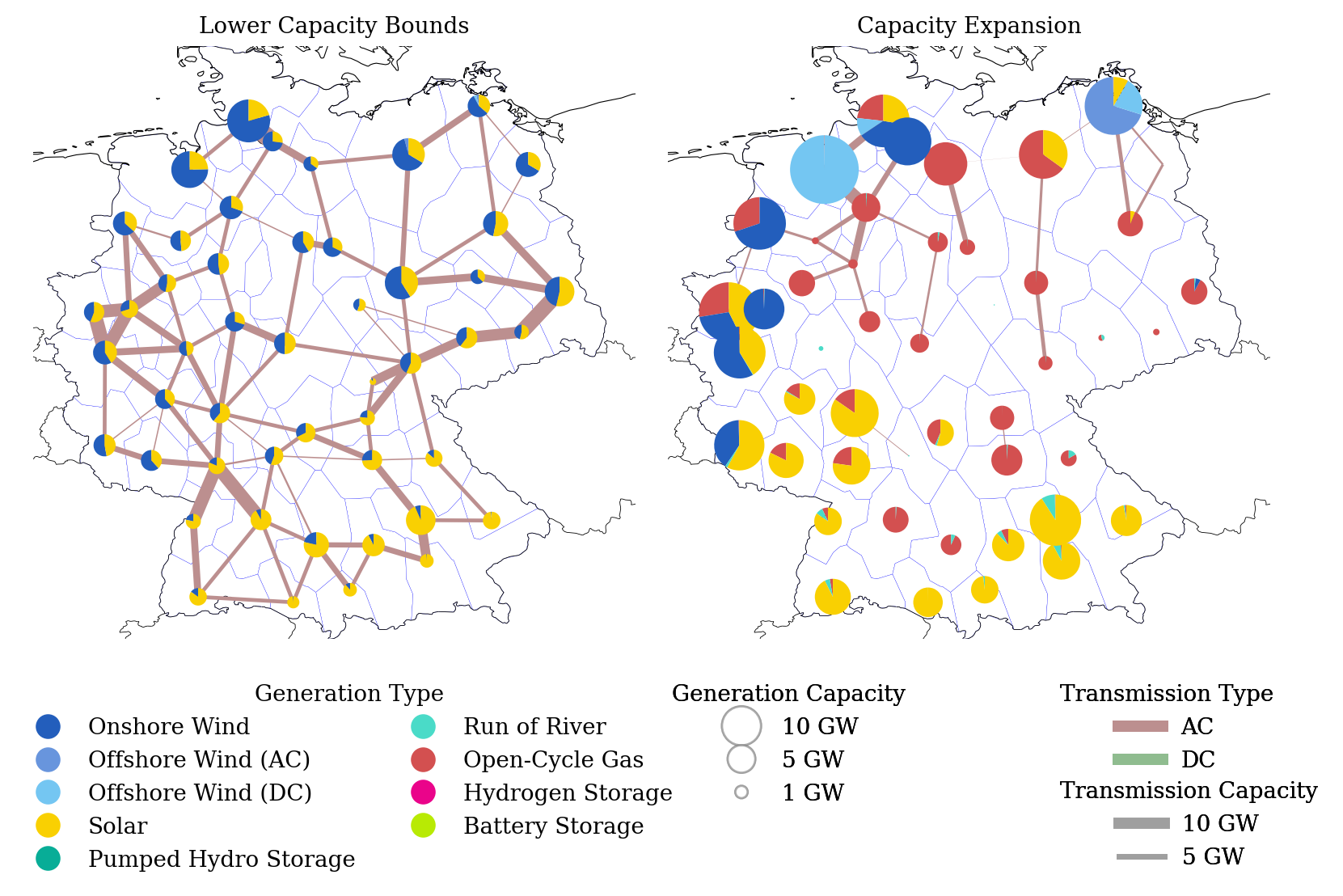}
    \caption{Model of the German power system. The left-hand side shows the minimum requirements for renewable and transmission capacities. The right-hand side shows the cost-optimal capacities of all generators as well as transmission expansions. The effective CO$_2$ price is set to 120 \euro\, per ton of CO$_2$ emission.}
    \label{fig:network}
\end{figure*}

\section{Application Cases}
\label{sec:application_case}

Using a cost-optimized model of the German power grid with 50 representative nodes and a time span of one year with hourly resolution, four use cases of the Price Tracing method are demonstrated.

The investment model builds on the PyPSA-EUR workflow \cite{horsch_jonas_pypsa-eur_2020} whose technical details and assumptions are presented in \cite{horsch_pypsa-eur_2018}. Transmission line capacities can be expanded but require a minimum of today's capacities, originally retrieved from the ENTSO-E Transmission System Map \cite{entso-e_entso-e_nodate}. Further, a minimum deployment of renewable capacity is required, derived from matching all entries of the OPSD renewable power plant list \cite{schlechtRenewablePowerPlants2020} to its closest bus. Wind and solar capacity expansion are limited by land-use restrictions considering agriculture, urban, forested and protected areas based on the CORINE and NATURA2000 database \cite{eea_corine_2012,eea_natura_2016}. \ac{PHS} and \ac{ROR} power plants are fixed to today's capacities and may not be expanded. In addition, unlimited expansion of batteries and H$_{2}$-storages and \ac{OCGT} are allowed at each node.
\begin{figure}
    \centering
    \includegraphics[width=\linewidth]{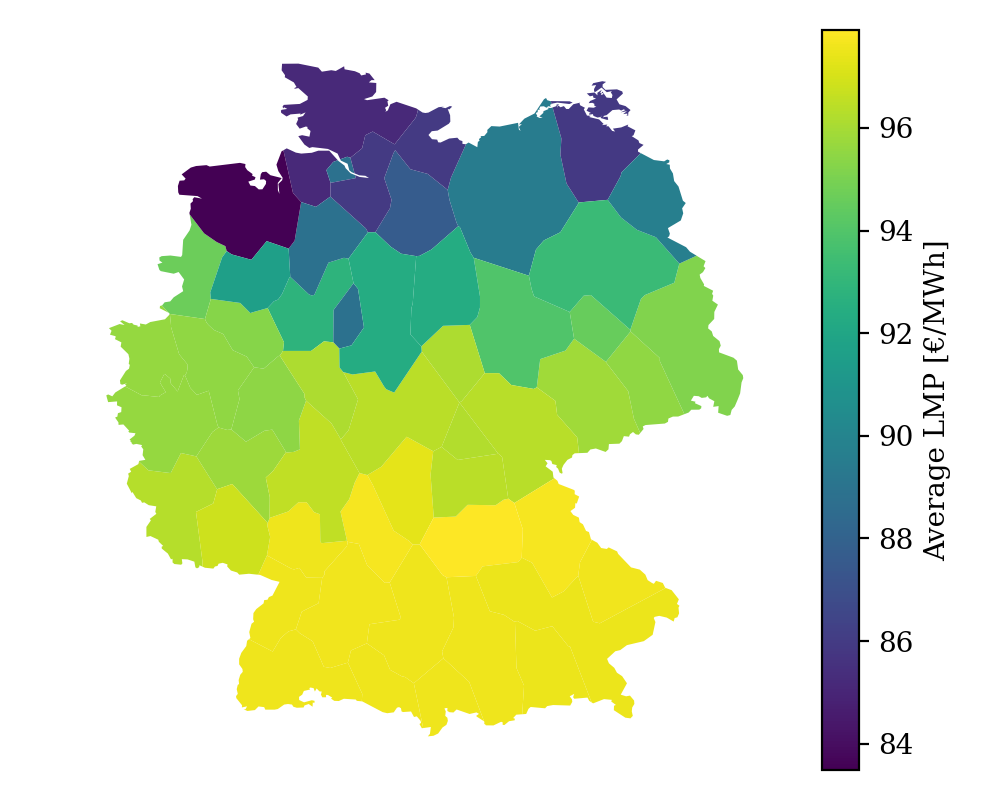}
    \caption{Load-weighted average electricity price $\averagelmp$ per region in the optimized German power system. Regions in the middle and south of Germany have high prices whereas electricity in the North with a strong wind, transmission and \ac{OCGT} infrastructure is cheaper.}
    \label{fig:average_price}
\end{figure}
We assume a carbon price of 120 \euro\, per ton of CO$_{2}$, which adds an effective price of 55 \euro/\megawatthour\, to the operational price of \acp{OCGT} (assuming a gross emission of 180~kg/MWh and an efficiency of 39\%). All cost assumptions on operational costs and annualized capital costs are summarized in \cref{tab:cost_assumptions}.

The network is shown in  \cref{fig:network}. The left-hand side shows all the minimum capacity requirements, the right-hand side the cost-optimal generation capacities as well as net transmission expansions. Solar capacities are expanded in the south, onshore and offshore wind in the upper north and far west. \ac{OCGT} are built in the middle and north of the country. Transmission lines are amplified along the north-south axis, including a major \ac{DC} link, associated with the German S\"ud-Link, running from the coastal region to the southwest.
The total annualized cost of the power system is roughly 46 billion \euro.

\Cref{fig:average_price} displays the load-weighted average electricity price $\averagelmp$ per region, defined by
\begin{align}
    \averagelmp = \dfrac{\sum_{t} \lmp \demand}{\sum_t \demand} = \dfrac{\sum_{t} \payment }{\sum_t \demand}
\end{align}
It reveals a relatively strong gradient from the south (at roughly 98 \euro/MWh) to the north (84 \euro/MWh). Regions with relatively poor renewable resources and therefore little capacity expansion tend to have higher prices.
\\

\subsection{Usage-based Network Tariff}

The first possible use case of the Price Tracing method discussed here is a usage-based network tariff. As in other countries, electricity consumers in Germany pay a uniform network tariff, the ``Netzentgelte''. These are based on a relatively opaque calculation and valuation of all network costs for each Transmission System Operator, in addition to an individually derived profit margin.

The Price Tracing method allocates all congestion revenues to consumers in the network, given by $\allocatelinecost$. This cost allocation can be used to define a usage-based network tariffs $\averagelmp^\text{grid}$ for each bus-region $n$ in the network:
\begin{align}
    \averagelmp^\text{grid} = \dfrac{\sum_{t,\ell} \allocatelinecost}{\sum_t \demand}.
\end{align}
This is the average price that consumers at $n$ pay for the transmission of electricity. These include compensation for all capital investments taken by the network operators. Consumers with a high degree of nodal autarky, i.e. with a little net electricity import, pay a lower network tariff and vice versa. This gives an incentive to regions with a sub-optimal deployment of renewable technologies to invest in local infrastructure. Note since the cost-optimum represents a long-term equilibrium, the total revenue per network asset equals the total \ac{OPEX} and \ac{CAPEX}. The network tariffs $\averagelmp^\text{grid}$ reflect this equilibrium and therefore do not result in any profits for the grid operator.

A further advantage of this network tariff allocation lies in the possibility of differentiating between network operators. The quantity
\begin{align}
    \averagelmp[n\rightarrow \ell]^\text{grid} = \dfrac{\sum_{t} \allocatelinecost}{\sum_t \demand}
\end{align}
decomposes the network tariff into single lines, naturally fulfilling  $\averagelmp^\text{grid} = \sum_\ell \averagelmp[n\rightarrow \ell]^\text{grid}$. On the one hand this enables operator or country specific network tariffs to be quantified so they are consistent with cross-border flows. On the other hand this facilitates tracing back which congested lines are the strongest drivers to the local network tariff.

\begin{figure}
    \includegraphics[width=\linewidth]{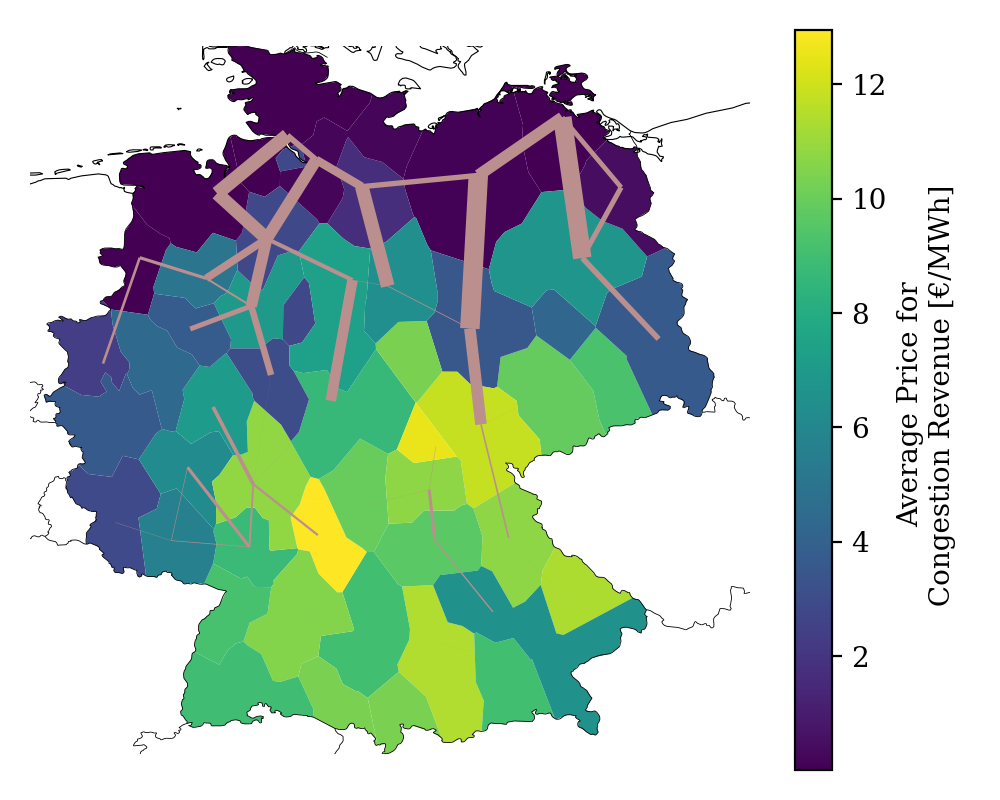}
    \caption{Usage-based network tariff per region. The resulting prices are indicated by the color of the region, the transmission lines are drawn in proportion to their congestion revenues.}
    \label{fig:congestion_revenue}
\end{figure}

\Cref{fig:congestion_revenue} shows the network tariffs $\averagelmp[n]^\text{grid}$ for all regions of the network model as well as the average congestion revenues per transmission line. Regions close to the shore hardly affect the transmission system, hence their network tariff is negligible. In contrast, regions in the middle and south are relying on power transfer from the north. Their demand is the main cause for the transmission system expansion. This translates to a higher network tariff.

\subsection{Emission Cost Allocation}
\label{sec:co2-cost-allocation}

\begin{figure}[h!]
    \includegraphics[width=\linewidth]{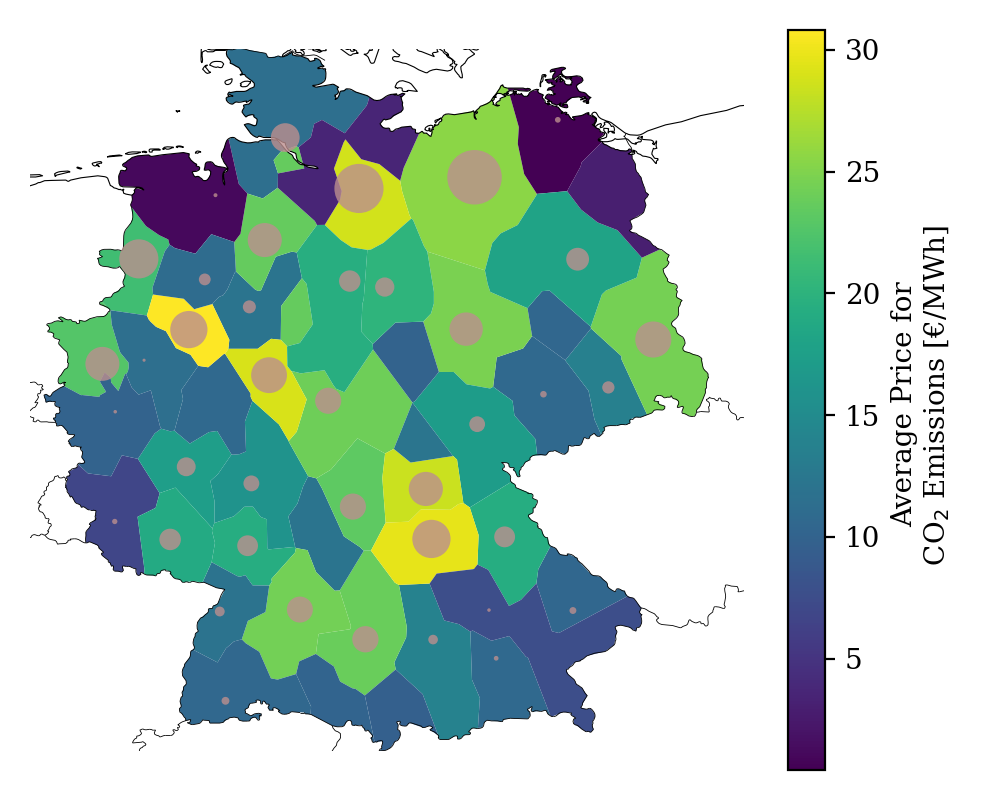}
    \caption{Average CO$_2$ cost per consumed MWh. The effective prices are indicated by the color of the region, the circles are drawn in proportion to the revenue per regional generators.}
    \label{fig:opex_price}
\end{figure}

The mapping $\allocategeneratorcost$ allocates all dispatch revenues to consumers. Naturally, the revenues of conventional generators include costs accounting for CO$_2$ emissions. Using the mechanism of the Price Tracing method, these CO$_2$ costs may be mapped to consumers by weighting the dispatch allocation $\allocategeneration$ with the effective CO$_2$ price $\emissionprice$ per produced MWh at generator $s$. Applying an emission cost allocation in the form of
\begin{align}
    \allocateemissioncost = \emissionprice \allocategeneration
\end{align}
encourages consumers to reduce their emission-intensive power consumption and leads to a transparent polluter pay principle. The average emission cost per consumed MWh is then given by
\begin{align}
    \averagelmp^{\text{CO}_2} = \dfrac{\sum_{t,\ell} \allocateemissioncost}{\sum_t \demand}.
\end{align}
\Cref{fig:opex_price} depicts the price for the network model. Regions with higher \ac{OCGT} backup capacity tend to have higher allocated emission costs. The size of the circles is proportional to the emission cost occasioned by the local generators.
Note, since the CO$_2$ price is given exogenously, this tracing method may also be applied to networks without a nodal pricing scheme.

\subsection{A Transparent Nodal Pricing Scheme}

\begin{figure*}[h!]
    \includegraphics[width=\linewidth]{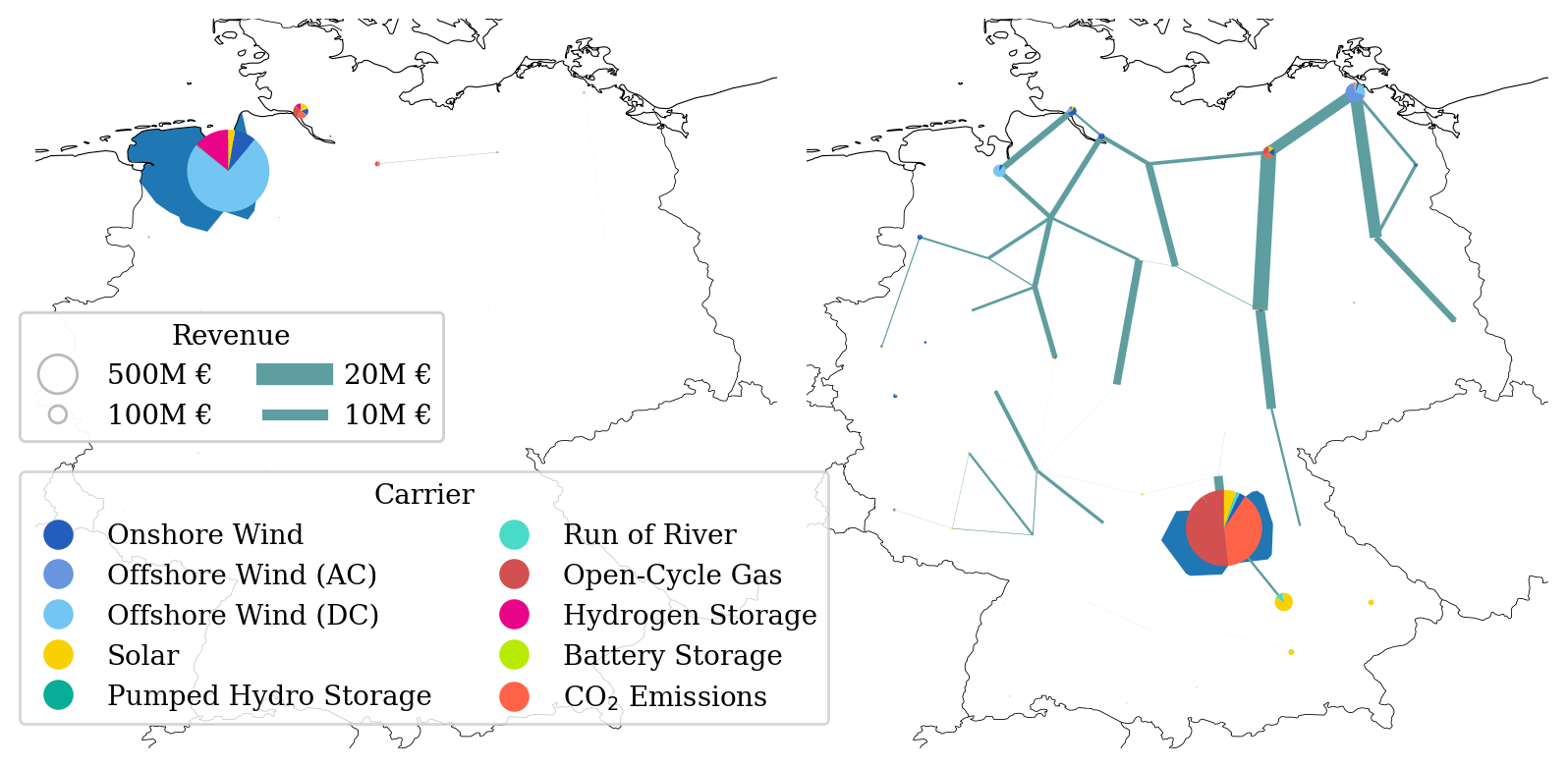}
    \caption{Comparison of region payments with the lowest \ac{LMP} (left) and the region with the highest \ac{LMP} (right). The region is colored in dark blue. The circles indicate to which bus and technology the payments are assigned. The thickness of the lines is proportional to dedicated payments.}
    \label{fig:direct-allocation}
\end{figure*}

Another application of the Price Tracing method is to make the nodal demand cost in a network more transparent. This answers the question of which parts of the system consumers' electricity payments go to. To demonstrate this, \cref{fig:direct-allocation} shows the total \ac{P2P} cost assignments $\sum_t \allocategeneratorcost$ and $\sum_t  \allocatelinecost$ for the region with the lowest average \ac{LMP} (left side) and for the region with the highest \ac{LMP} (right side). The dedicated region is indicated in blue. The circles and their segments indicate the allocated dispatch revenues which are subdivided by carrier type. The lines indicate the allocated congestion revenue.

The low-cost region in the Northwest has few congestion costs due to its low net imports. It profits from local offshore wind farms and only a small share of the payments is allocated to remote \acp{OCGT}. In contrast, the high-price region is highly dependent on local \ac{OCGT} and the transmission system, which causes high amounts of allocated dispatch and congestion costs. Interestingly, its payments to  onshore and offshore wind infrastructure are low despite a third of its supply coming from wind power, compare \cref{fig:power_mix}. This suggests that wind power supply in this region is constrained not by depleted wind power resources, but by bottlenecks in the transmission grid.



\subsection{Revenue Decomposition}

Finally, a more abstract use case of Price Tracing is demonstrated that focuses on the decomposition of congestion and dispatch revenues for each asset. In a long-term equilibrium without any exogenous constraints, a network asset recovers all its \ac{OPEX} and \ac{CAPEX} from the revenue. This relation is known as the zero-profit rule and is an outcome of the \ac{KKT} conditions of the underlying cost-minimization problem, derived in detail in \cref{sec:zero_profit_generation,sec:zero_profit_flow,sec:zero_profit_storage_units}. However, if the cost-optimum is subject to additional constraints, for example by a minimum share of renewable capacity, the zero-profit relation is altered and more cost contributions have to be considered \cite{brown_decreasing_2020} as soon as the new constraint becomes binding. With the Price Tracing method it is possible to decompose the effect of the additional boundary conditions to the cost of individual consumers.

\begin{table}[h!]
    \caption{Exemplary contributions to the revenue of network assets highlighted in this work. Depending on the formulation of the optimization problem, the list changes and may possibly include other terms.}
    \begin{center}
        \begin{tabular}{c|c p{0.3\textwidth}}
            \toprule
            \text{Contribution}  & Symbol          \\
            \midrule
            \text{OPEX}          & $\opex$         \\
            \text{CAPEX}         & $\capex$        \\
            \text{Emission Tax}  & $\emissioncost$ \\
            \text{Scarcity Rent} & $\scarcitycost$ \\
            \text{Subsidies}     & $\subsidycost$  \\
            \vdots               & \vdots          \\
            \bottomrule
        \end{tabular}
    \end{center}
    \label{tab:contributions}
\end{table}

\noindent
\Cref{tab:contributions} shows a possible set of contributions to the asset revenue. On top of \ac{OPEX} and \ac{CAPEX}, there are possible emission costs which were previously discussed in \cref{sec:co2-cost-allocation}. In presence of capacity expansion limits, additional Scarcity Rent $\scarcitycost$ must be included in the revenues, in case the corresponding constraints are binding, for mathematical details see \cref{sec:upper_capacity_limits}. This rent translates to a compensation for higher investments in a highly competitive market, but can also be ignored in case an asset is fixed to its existent, amortized capacity. Likewise, assets may be constrained to lower capacity requirements which reflect \ie  existing infrastructure, called Brownfield constraints, see \cref{sec:lower_capacity_limits}. If binding, the lower limit reduces the revenue of the corresponding asset and leaves a Subsidy Cost $\subsidycost$ which have to be compensated by external institutions (government, community) or are simply ignored when the asset is already amortized.

Except for the subsidies, all contributions can be expressed in terms of the operation of an asset, that is the generation $\generation$ and the flow $\flow$. Thus, given the set of discussed revenue contributions, the dispatch revenue allocation decomposes into
\begin{align}
    \allocategeneratorcost = \allocateopex + \allocatecapex + \allocateemissioncost + \allocatescarcitycost - \subsidycost_s .
\end{align}
where all expressions are summarized in \cref{tab:cost_allocation_map}. \Cref{sec:application_case_appendix} shows the individual cost allocations for the network model. Note that transmission lines are assumed to have neither \ac{OPEX} nor emission tax, \ie
\begin{align}
    \allocatelinecost = \capex_{n \ra \ell, t} + \scarcitycost_{n \ra \ell, t} - \subsidycost_\ell
\end{align}
This approach allows modelers and system operators to disentangle the combined effect of multiple constraints on the dispatch and congestion revenues. For example, when analyzing these cost allocations for the presented German network model, it becomes clear that buses, which are far from wind farms, pay higher prices due to increased reliance on transmission and backup capacity. On the other hand, buses with high renewable installation spend most payments to local assets, see \cref{sec:application_case_appendix} for details.

\section{Limitations}
\label{sec:limitations}

The method presented is closely linked to the underlying power system model and its optimization. In the usual electricity market mechanism, electricity producers bid on the market and Transmission System Operators make profits by charging grid fees. On the other hand, electricity consumers always pay indirectly for electricity generation costs, grid usage fees, and carbon emission costs.
While this study is less concerned with today's market structures, it is more concerned with showing what other market structures are theoretically possible and what flexibilities result from them. Nodal price structures with optimized prices hold many possibilities to create incentives. The preceding explanation is intended to help illuminate this space of possibilities.

The presented cost allocation is based on the linear power flow approximation. Yet, the method is equally applicable to a system with an optimal power flow, \ie an \ac{AC} power flow. However, in this case, the allocations $\allocategeneration$ and $\allocateflow$ should be computed with the Z-Bus flow allocation presented in \cite{conejo_z-bus_2007} which, by design, respects both Kirchhoff Circuit Laws. Here, an additional cost term accounting for the transmission loss has to be considered in the decomposition of the congestion revenue.

The used optimization does not take security constraints of the transmission system into account. These may be incorporated following the approach in \cite{nikoukar_transmission_2012}.

The method can be equally applied to short and long-term planning network models. However, the paper does not take the detailed structure of today's power markets into account. These vary from country to country and often reveal interlocked mechanisms (energy only markets, reserve markets, redispatch etc.). The aim of the presented work is to reconsider some of the mechanisms and  to reevaluate their market efficiencies.

\section{Conclusion}
\label{sec:conclusions}

A new method, called Price Tracing, was presented. It builds upon the nodal pricing scheme and decomposes the dispatch and congestion revenues of generators and transmission lines into individual payments of consumers. It is based on the Average Participation method which calculates assignments between generators and consumers using the principle of proportional sharing. These assignments are weighted by the market price to determine the allocation of the revenues to consumers. By means of a numerical example, it was shown that there exist two options of how to treat network cycles regarding cost allocation and laid out why it is more sensible to consider shadow prices of the Kirchhoff Voltage constraint.

Using an optimized model of the German power system with a price of 120~\euro\, per ton of CO$_2$, four use cases of the Price Tracing method were presented. First, it was demonstrated how grid tariffs can be adjusted to a transparent, consumption-based tariff that incentivizes investment in local infrastructure. Second, Price Tracing allows the grid to track the costs associated with CO$_2$ emissions. These results can be used to reduce CO$_2$-intensive demand on a grid. The third use case presented uses Price Tracing to derive cost contributions for individual consumers to make nodal payments transparent and explain price differences. This relates to the fourth case where individual dispatch and congestion revenues are decomposed into contributions from operational expenditures, capital expenditures, scarcity rents, and subsidies. It was shown that the allocation can be performed for a single cost contribution to provide insights into electricity system models or to derive policy advice.

\subsubsection*{Reproducibility and Expansion}

All figures and data points can be reproduced by using the \textit{python} workflow in \cite{hofmann_pypsa-costallocation_2020}. The automated workflow allows for higher spatial resolution of the network (scalable up to a the full ENTSO-E Transmission System Map) and optionally taking the total European power system into account.

\subsection*{Funding}
This research was funded by the by the German Federal Ministry for Economics Affairs and Energy (BMWi) in
the frame of the NetAllok project (grant number 03ET4046A) \cite{bundesministeriumfurwirtschaftundenergieVerbundvorhabenNETALLOKMethoden}.

\subsubsection*{Acknowledgement}

In particular, we thank Tom Brown for very fruitful discussions. We are very grateful to Alexander Zerrahn for reviewing. Further, we thank Alexander Kies and Horst Stöcker for their creative and fruitful inputs.


\clearpage
\appendix
\subsection*{Acronyms}
\begin{acronym}[UMLX]
    \acro{AC}{Alternating Current}
    \acro{AP}{Average Participation}
    \acro{CAPEX}{Capital Expenditures}
    \acro{CO2}[CO$_2$]{Carbon Dioxide}
    \acro{DC}{Direct Current}
    \acro{FA}{Flow Allocation}
    \acro{HVAC}{High-Voltage Alternating Current}
    \acro{HVDC}{High-Voltage Direct Current}
    \acro{Hydro}{Hydro-Electric Dams}
    \acro{Hydrogen}[H$_2$-storage]{Hydrogen Storage}
    \acro{KCL}{Kirchhoff's Current Law}
    \acro{KKT}{Karush–Kuhn–Tucker}
    \acro{KVL}{Kirchhoff's Voltage Law}
    \acro{LCOE}{Levelized Cost of Electricity}
    \acro{LMP}{Locational Marginal Price}
    \acro{LOPF}{Linear Optimal Power Flow}
    \acro{OCGT}{Open-Cycle Gas Turbine}
    \acro{OPEX}{Operational Expenditures}
    \acro{P2P}{Peer-to-Peer}
    \acro{PHS}{Pumped-Hydro-Storage}
    \acro{PTDF}{Power Transfer Distribution Factors}
    \acro{PV}{Photovoltaic}
    \acro{PyPSA-EUR}{PyPSA-Europe}
    \acro{PyPSA-Eur}{\acs*{PyPSA} Europe}
    \acro{PyPSA}{Python for Power System Analysis}
    \acro{ROR}{Run-of-River}
    \acro{SOAF}[SO\&AF]{\acs*{ENTSO-E} Scenario Outlook and Adequacy Forecast}
    \acro{TSO}{Transmission System Operator}
    \acrodefplural{TSO}{Transmission System Operators}
\end{acronym}

\renewcommand\theequation{\thesection.\arabic{equation}}
\setcounter{equation}{0}

\renewcommand\thefigure{\thesection.\arabic{figure}}
\setcounter{figure}{0}

\section{Network Optimization}

\subsection{Optimality Conditions}
\label{sec:optimality_conditions}

Consider the following linear minimization problem

\begin{align}
    \min_{x_n} \sum_n c_n \, x_n
\end{align}
subject to inequality constraints
\begin{align}
    g_i(x_n) \le 0 \resultsin{\mu_i}
\end{align}
and equality constraints
\begin{align}
    h_j(x_n) = 0 \resultsin{\lambda_j}
\end{align}
where $\mu_i$ and $\lambda_j$ are the corresponding dual variables, also known as \ac{KKT} variables.
The Lagrangian is given by
\begin{align}
    \lagrangian = \sum_n c_n x_n + \sum_i \mu_i g_i(x_n) + \sum_j \lambda_j h_j(x_n).
\end{align}

At the optimum $x^*_n$ the following \ac{KKT} conditions are satisfied.
\begin{enumerate}
    \item Stationarity \\
          \begin{align}
              \pdv{\lagrangian}{x_n} = c_n + \sum_i \mu_i \pdv{g_i}{x_n} + \sum_j \lambda_j \pdv{h_j}{x_n} & = 0
          \end{align}
    \item Primal Feasibility \\
          \begin{align}
              g_i(x_n) \le 0 \qquad \forall i \\
              h_j(x_n) \le 0 \qquad \forall j
          \end{align}
    \item Dual Feasibility \\
          \begin{align}
              \mu_i \ge 0 \qquad \forall i
          \end{align}
    \item Complementary Slackness \\
          \begin{align}
              \sum_i \mu_i g_i(x_n) = 0
          \end{align}
\end{enumerate}

\subsection{Problem formulation}
\label{sec:problem_formulation}

We follow a cost-minimization approach, which optimizes all operational and capital expenditures of the network. The corresponding objective is given by

\begin{align}
    \min_{g_{s,t}, G_s, F_{\ell, t}} \sum_{s,t} o_s g_{s,t} + \sum_s c_s G_s + \sum_\ell c_\ell F_\ell
\end{align}
where $G_s$ and $F_\ell$ are the nominal capacities per generator $s$ and line $\ell$ respectively. The operational costs of generators are given by $o_s$, the capital costs for generators and lines by $c_s$ and $c_\ell$.

Of particular importance is the nodal balance constraint which ensures that the amount of power that flows into a bus equals the power that flows out of a bus, thus reflects the \ac{KCL}. With a given demand $\demand$ this translates to
\begin{align}
    \nodalgeneration - \sum_\ell \incidence \flow & =  \demand
    \resultsin{\lmp} \Forall{n,t}
    \label[constraint]{eq:nodal_balance_lin}
\end{align}
where $\incidence$ is +1 if line $\ell$ starts at bus $n$, -1 if it ends at $n$, 0 otherwise. The nodal generation $\nodalgeneration$ collects the production of all nodal assets. The corresponding dual variable (or shadow price) $\lmp$ is the \ac{LMP} per bus and time step. In an economic equilibrium for a nodal pricing scheme this is the \euro/\megawatthour\, price which a consumer has to pay.\\

In order to impose the \ac{KVL} for the linearized \ac{AC} flow, the constraint
\begin{align}
    \sum_{\ell} \cycle \, \reactance \, \flow = 0 \resultsin{\cycleprice} \Forall{c,t}
\end{align}
is added to the problem. Here, $\reactance$ denotes the line's impedance and $\cycle$ is +1 if $\ell$ lays aligned to the network cycle $c$, --1 if it lays in the opposite direction and zero if it is not part of the network cycle $c$.
For simplicity we define the total price of the \ac{KVL} per line as
\begin{align}
    \lmpkvl = \sum_c \cycleprice\, \cycle\, \reactance \Forall{\ell,t} .
\end{align}
The total price per transmitted unit of power can now be defined by the price differences of the starting and ending bus of the transmission line
\begin{align}
    \lmp[\ell] = \lmpdiff
\end{align}
with
\begin{align}
    \lmpdiff = -\sum_n \incidence \lmp .
\end{align}
or it additionally includes the price for the \ac{KVL}
\begin{align}
    \lmp[\ell] = \lmpkvl + \lmpdiff .
    \label{eq:congestion_price}
\end{align}

\subsection{Zero Profit Generation}
\label{sec:zero_profit_generation}

For each generator $s$ the optimization defines a lower and upper limit for the power output $\generation$, given by
\begin{align}
    - \generation                     & \le 0 \resultsin{\mulowergeneration} \Forall{s,t}
    \label[constraint]{eq:lower_generation_capacity_constraint}                           \\
    \generation - \capacitygeneration & \le 0 \resultsin{\muuppergeneration} \Forall{s,t}
    \label[constraint]{eq:upper_generation_capacity_constraint}
\end{align}
\Cref{eq:upper_generation_capacity_constraint,eq:lower_generation_capacity_constraint}, which yield the \ac{KKT} variables $\muuppergeneration$ and $\mulowergeneration$, imply the complementary slackness,
\begin{align}
    \muuppergeneration \left( \generation - \generationpotential \, \capacitygeneration \right) & = 0  \Forall{s,t}
    \label{eq:complementary_slackness_upper_generation}                                                             \\
    \mulowergeneration  \, \generation                                                          & = 0 \Forall{s,t}
    \label{eq:complementary_slackness_lower_generation}
\end{align}

The stationarity of the generation capacity variable leads to
\begin{align}
    \pdv{\lagrangian}{\capacitygeneration}  = 0 \,\, \rightarrow \,\,
    \capitalpricegeneration =  \sum_t \muuppergeneration \, \generationpotential  \Forall{s}
    \label{eq:capex_generation_duality}
\end{align}
and the stationarity of the generation to
\begin{align}
    \pdv{\lagrangian}{\generation} & = 0 \,\, \rightarrow \,\,
    \operationalpricegeneration =  \sum_n \incidencegenerator \, \lmp - \muuppergeneration + \mulowergeneration \Forall{s} \label{eq:opex_duality}
\end{align}
where $\incidencegenerator$ is set to one if generator $s$ is placed at node $n$ and zero otherwise.

Multiplying both sides of \cref{eq:capex_generation_duality} with $\capacitygeneration$ and using \cref{eq:complementary_slackness_upper_generation} leads to
\begin{align}
    \capitalpricegeneration \, \capacitygeneration  = \sum_t \muuppergeneration \, \generation \Forall{s}
    \label{eq:capital_price_generation_sum}
\end{align}
The zero-profit rule for generators is obtained by multiplying \cref{eq:opex_duality} with $\generation$ and using \cref{eq:complementary_slackness_lower_generation,eq:capital_price_generation_sum} which results in
\begin{align}
    \capitalpricegeneration \, \capacitygeneration + \sum_t \operationalpricegeneration \generation = \sum_{n,t} \lmp \incidencegenerator \generation \Forall{s}
\end{align}
It states that over the whole time span, all \ac{OPEX} and \ac{CAPEX} for generator $s$ (left-hand side) are payed back by its revenue (right-hand side).

\subsection{Zero Profit Transmission System}
\label{sec:zero_profit_flow}

For transmission lines the flow $\flow$ is limited by the nominal capacity $\capacityflow$ in both directions, mathematically translating to
\begin{align}
    \flow - \capacityflow   & \le 0 \resultsin{\muupperflow} \Forall{\ell,t}
    \label[constraint]{eq:upper_flow_capacity_constraint}                      \\
    - \flow - \capacityflow & \le 0 \resultsin{\mulowerflow} \Forall{\ell,t} .
    \label[constraint]{eq:lower_flow_capacity_constraint}
\end{align}
The \ac{KKT} variables $\muupperflow$ and $\mulowerflow$ are only non-zero if $\flow$ is limited by the transmission capacity in positive or negative direction, i.e. \cref{eq:upper_flow_capacity_constraint} or \cref{eq:lower_flow_capacity_constraint} are binding. For flows below the thermal limit, the complementary slackness
\begin{align}
    \muupperflow \left( \flow - \capacityflow \right) & = 0 \Forall{\ell,t}
    \label{eq:complementary_slackness_upper_flow}                            \\
    \mulowerflow \left( \flow - \capacityflow \right) & =  0 \Forall{\ell,t}
    \label{eq:complementary_slackness_lower_flow}
\end{align}
sets the respective \ac{KKT} variables to zero.

The stationarity of the transmission capacity leads to
\begin{align}
    \pdv{\lagrangian}{\capacityflow} = 0 \,\, \rightarrow \,\,
    \capitalpriceflow =  \sum_t \left( \muupperflow - \mulowerflow \right) \Forall{\ell}
    \label{eq:capacity_flow_duality}
\end{align}
and the stationarity with respect to the flow to
\begin{align}
    0 & = \pdv{\lagrangian}{\flow}                                                                                                                 \\
    0 & = - \sum_n \incidence \lmp  + \sum_c \cycleprice \cycle \reactance  - \muupperflow + \mulowerflow \Forall{\ell, t} \label{eq:flow_duality}
\end{align}

When multiplying \cref{eq:capacity_flow_duality} with $\capacityflow$ and using the complementary slackness \cref{eq:complementary_slackness_upper_flow,eq:complementary_slackness_lower_flow} we obtain
\begin{align}
    \capitalpriceflow \, \capacityflow = \sum_t \left( \muupperflow - \mulowerflow \right)  \, \flow \Forall{\ell}
    \label{eq:capital_price_flow_sum}
\end{align}
Again we can use this to formulate the zero-profit rule for transmission lines. We multiply \cref{eq:flow_duality} with $\flow$, which finally leads us to
\begin{align}
    \capitalpriceflow \, \capacityflow & = - \sum_n \incidence\, \lmp\, \flow + \sum_c \cycleprice\, \cycle\, \reactance\, \flow
    \Forall{\ell}                                                                                                                \\
                                       & = \sum_t \lmp[\ell] \flow \Forall{\ell}
\end{align}
where $\lmp[\ell]$ is given by \cref{eq:congestion_price}.
It states that the congestion revenue of a line (first term right-hand side) reduced by the cost for cycle constraint exactly matches its \ac{CAPEX}.

\subsection{Zero Profit Storage Units}
\label{sec:zero_profit_storage_units}


For an simplified storage model, the upper capacity $\capacitystorage$ limits the discharging dispatch $\storagedispatch$, the storing power $\storagecharge$ and state of charge $\storagesoc$ of a storage unit $r$ by
\begin{align}
    \storagedispatch - \capacitystorage   & \le 0  \resultsin{\muupperstoragedispatch} \Forall{r,t} \\
    \storagecharge - \capacitystorage     & \le 0  \resultsin{\muupperstoragecharge} \Forall{r,t}   \\
    \storagesoc - h_r \, \capacitystorage & \le 0  \resultsin{\muupperstoragesoc} \Forall{r,t}
\end{align}
where we assume a fixed ratio between dispatch and storage capacity of $h_r$.
The state of charge must be consistent throughout every time step according to what is dispatched and stored,
\begin{align}
    \notag
    \storagesoc - \efficiencysoc \storageprevioussoc - \efficiencycharge \storagecharge & + (\efficiencydispatch)^{-1} \storagedispatch = 0 \\
                                                                                        & \resultsin{\mustateofcharge} \Forall{r,t}
\end{align}

We use the result of Appendix B.3 in \cite{brown_decreasing_2020} which shows that a storage recovers its capital (and operational) costs from aligning dispatch and charging to the LMP, thus
\begin{align}
    \notag
    \sum_t \operationalpricestorage \, \storagedispatch + \capitalpricestorage \, \capacitystorage = \sum_t \lmp \incidencestorage & \left(\storagedispatch - \storagecharge \right) \Forall{r,t}
\end{align}
where $\incidencestorage$ is one if storage $r$ is paced at node $n$ and zero otherwise.
The stationarity of the dispatched power leads us to
\begin{align}
    \notag
    \pdv{\lagrangian}{\storagedispatch}                                                                                                                            & = 0                   \\
    \operationalpricestorage - \sum_n \lmp \, \incidencestorage - \mulowerstoragedispatch + \muupperstoragedispatch + (\efficiencydispatch )^{-1} \mustateofcharge & = 0 \;  \forall\; r,t
    \label{eq:stationarity_storagedispatch}
\end{align}
which we  can use to define the revenue which recovers the \ac{CAPEX} at $r$,
\begin{align}
    \notag
    \capitalpricestorage \, \capacitystorage = \sum_t \left(\muupperstoragedispatch - \mulowerstoragedispatch  + (\efficiencydispatch )^{-1} \mustateofcharge \right) \storagedispatch \\
    - \sum_t \lmp \incidencestorage  \storagecharge \Forall{r}
    \label{eq:no_profit_capex_storage2}
\end{align}


It stands to reason to assume that when a storage charges power, it does not supply any demand. Thus consumers only pay storage units in times the storage dispatches power. Hence, we restrict the allocatable revenue per storage unit to the first term in \cref{eq:no_profit_capex_storage2}. This allocates the CAPEX of $r$ plus the costs $\remainingcost^E_r$ it needs to buy the charging power,
\begin{align}
    \capexstorage_r + \remainingcost^E_r = \sum_t \left(\muupperstoragedispatch - \mulowerstoragedispatch  + (\efficiencydispatch )^{-1} \mustateofcharge \right) \storagedispatch
\end{align}
In charging times the total of remaining costs $\remainingcost^E_r$ is spent to power from other assets. These costs scale with the amount of installed storage capacity.

\subsection{Upper Capacity Expansion Limits}
\label{sec:upper_capacity_limits}

\begin{subequations}
    In real-world setups the capacity expansion of generators, lines or other assets are often limited. This might be due to land use restrictions or social acceptance considerations.
    When constraining the capacity $\capacity$ of an arbitrary asset $i$ ($i$ may now be any generator, line etc.) to an upper limit $\capacityupper$, in the form of
    \begin{align}
        \capacity - \capacityupper \le 0 \resultsin{\muuppernom}
        \label{eq:capacityexpansionmaximum},
    \end{align}
    the zero profit condition alters as soon as the constraint becomes binding. Then, asset $i$ is payed an additional scarcity rent
    \begin{align}
        \scarcitycost_i = - \muuppernom \capacity \Forall{i \in I}
        \label{eq:scarcitycost}
    \end{align}
    This rent may account for different possible realms, as for example the increased market price in higher competed areas or additional costs for social or environmental compensation. The share in $\allocatecost$ which consumers pay for the scarcity rent can be recalculated by a correct weighting of the shadow price $\muuppergenerationnom$ with the capital price $c_i$, leading to
    \begin{align}
        \allocatescarcitycost = \dfrac{\muuppernom}{c_i + \muuppernom} \, \allocatecost \Forall{i}
    \end{align}
\end{subequations}

\subsection{Lower Capacity Expansion Limits}
\label{sec:lower_capacity_limits}

\begin{subequations}
    In order to take already built infrastructure into account, the capacity $\capacity$ of an arbitrary asset $i$ may be constrained by a minimum required capacity $\capacitylower_i$. This introduces a constraint of the form
    \begin{align}
        \capacitylower - \capacity  \le 0 \resultsin{\mulowernom}
        \label{eq:capacityexpansionminimum}
    \end{align}
    Again, such a setup alters the zero profit condition of asset $i$, as soon as the constraint becomes binding.
    In that case, asset $i$ does not collect enough revenue from $\allocatecost$ in order to compensate the \ac{CAPEX}. The difference, given by
    \begin{align}
        \subsidycost_i = \mulowernom \capacity
        \Forall{i}
    \end{align}
    has to be subsidized by governments or communities or is simply ignored when investments are amortized. Allocating the subsidies to consumers is ambiguous as revenue of the assets are not in equilibrium with their \ac{CAPEX} and \ac{OPEX} anymore.
\end{subequations}

\subsection{Proof: Equivalence of local and imported prices}
\label{sec:proof_equivalence}

When inserting
\begin{align}
    \allocateflow = \sum_m \ptdf[m] (\allocatepeer - \delta_{nm} \demand)
    \label{eq:allocateflow2}
\end{align}
into
\begin{align}
    \lmp\, \demand & = \sum_m \lmp[m] \allocatepeer + \sum_\ell \lmp[\ell] \allocateflow \Forall{n,t}
    \label{eq:demand-cost-allocation2}
\end{align}
where
\begin{align}
    \lmp[\ell] = - \sum_m \incidence[m] \lmp[m]  + \sum_c  \cycleprice \cycle \reactance
\end{align}
the second expression on the right-hand side of \cref{eq:demand-cost-allocation2} becomes

\begin{align}
    \notag
      & - \sum_{\ell, m'} \ptdf[m'] \left( \allocatepeer[m' \rightarrow n]  - \delta_{n m'} \demand \right) \sum_m \incidence[m] \lmp[m]          \\
    \notag
      & + \sum_{\ell, m'} \ptdf[m'] \left( \allocatepeer[m' \rightarrow n]  - \delta_{n m'} \demand \right) \sum_c  \cycleprice \cycle \reactance \\
    \notag
    = & - \sum_{\ell, m', m} \ptdf[m'] \left( \allocatepeer[m' \rightarrow n]  - \delta_{n m'} \demand \right) \incidence[m] \lmp[m]              \\
    \notag
    = & - \sum_{m', m} \delta_{m m'} \left( \allocatepeer[m' \rightarrow n]  - \delta_{n m'} \demand \right) \lmp[m]                              \\
    \notag
    = & - \sum_{m} \left( \allocatepeer - \delta_{n m} \demand \right) \lmp[m]                                                                    \\
    = & - \sum_{m} \allocatepeer \lmp[m] + \demand \lmp
\end{align}
where in the third step we used the relation $\sum_\ell \ptdf \incidence[m] = \delta_{n m}$.
The second term in the first line vanishes as the basis cycles $\cycle$ are the kernel of the PTDF, $\sum_{\ell} \cycle \ptdf  = 0 \; \forall\, c, n$.

\begin{table*}[t]
    \begin{center}
        \begin{tabular}{l|c|c|c|c|c}
                               & Symbol             & Total Cost                                              & Allocation                                                                                                                               \\
            \toprule
            OPEX Production    & $\opexgeneration$  & $\sum_{t} \operationalpricegeneration \, \generation$   & $\operationalpricegeneration \allocategeneration $                                                                                       \\
            OPEX Storage       & $\opexstorage$     & $\sum_{t} \operationalpricestorage \, \storagedispatch$ & $\operationalpricestorage \allocatestoragedispatch $                                                                                     \\
            \midrule
            Emission Cost      & $\emissioncost$    & $ \emissionprice \, \emission \, \generation$           & $\emissionprice \,\emission \allocategeneration$                                                                                         \\
            \midrule
            CAPEX Production   & $\capexgeneration$ & $ \capitalpricegeneration \capacitygeneration$          & $\muuppergeneration \allocategeneration$                                                                                                 \\
            CAPEX Transmission & $\capexflow$       & $ \capitalpriceflow \capacityflow$                      & $\left(\muupperflow  \mulowerflow \right) \allocateflow$                                                                                 \\
            CAPEX Storage      & $\capexstorage$    & $ \capitalpricestorage \capacitystorage$                & $ \left(\muupperstoragedispatch  \mulowerstoragedispatch  + (\efficiencydispatch )^{1} \mustateofcharge\right) \allocatestoragedispatch$ \\
        \end{tabular}
    \end{center}
    \caption{Different cost terms and cost allocations of different assets, index $s$ refers to generators, $\ell$ to lines and $r$ to storage units. The definition of the cost weightings are defined in \cref{sec:zero_profit_generation,sec:zero_profit_flow,sec:zero_profit_storage_units}.}
    \label{tab:cost_allocation_map}
\end{table*}

\section{Power Allocation}
\label{sec:net_ap}

\newcommand{\incidenceM}{K}
\newcommand{\flowM}{f}
\newcommand{\injectionM}{p}
\newcommand{\slackM}{k}
\newcommand{\DirectedIncidence}{\mathcal{K}}
\newcommand{\InverseAPInjection}{\mathcal{J}}
\newcommand\diag[1]{\operatorname{diag}\left(#1\right)}

Allocating net injections using the AP method is derived from \cite{achayuthakan_electricity_2010}. In a lossless network the downstream and upstream formulations result in the same P2P allocation which is why we restrict ourselves to the downstream formulation only. In a first step we define a time-dependent auxiliary matrix $\InverseAPInjection_t$ which is the inverse of the $N\times N$ with directed power flow $m \rightarrow n$ at entry $(m, n)$ for $m \ne n$ and the total flow passing node $m$ at entry $\left( m, m\right)$ at time step $t$. Mathematically this translates to

\begin{align}
    \InverseAPInjection_t = \left( \diag{\injectionM^+} + \DirectedIncidence^- \diag{\flowM} \, \incidenceM \right)_t^{-1}
\end{align}
where $\DirectedIncidence^-$ is the negative part of the directed Incidence matrix $\DirectedIncidence_{n,\ell} = \text{sign}\left( f_\ell \right)  \incidence$. Then the P2P allocation for time step $t$ is given by
\begin{align}
    \allocatepeer = \InverseAPInjection_{m,n,t} \, \netproduction[m] \, \netconsumption
\end{align}

\newpage
\section{Application Case}
\label{sec:application_case_appendix}

\begin{table*}[h]
    \centering
    \input{tables/de50/prices}
    \caption{Operational and capital price assumptions for all type of assets used in the working example. The capital price for transmission lines are given in [k\,\euro/MW/km]. The cost assumptions are retrieved from the PyPSA-EUR model \cite{horsch_jonas_pypsa-eur_2020}.}
    \label{tab:cost_assumptions}
\end{table*}

The following figures contain more detailed information about the peer-to-peer cost allocation discussed in \cref{sec:application_case}. The cost or prices payed by consumers are indicated by the region color. The dedicated revenue is displayed in proportion to the size of circles (for assets attached to buses) or to the thickness of transmission branches.

\begin{figure}[h]
    \includegraphics[width=\linewidth]{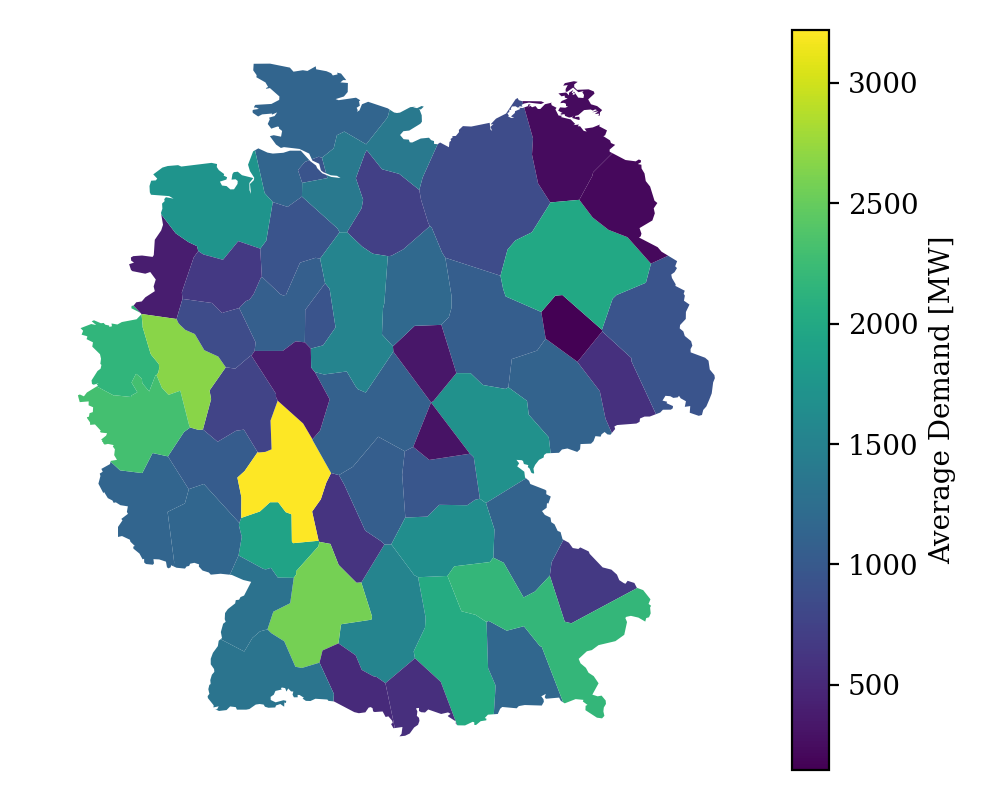}
    \caption{Average demand, $\sum_t \demand/T$ per regions. The regions with high population densities and larger areas reveal a higher demand.}
    \label{fig:average_demand}
\end{figure}

\begin{figure}
    \includegraphics[width=\linewidth]{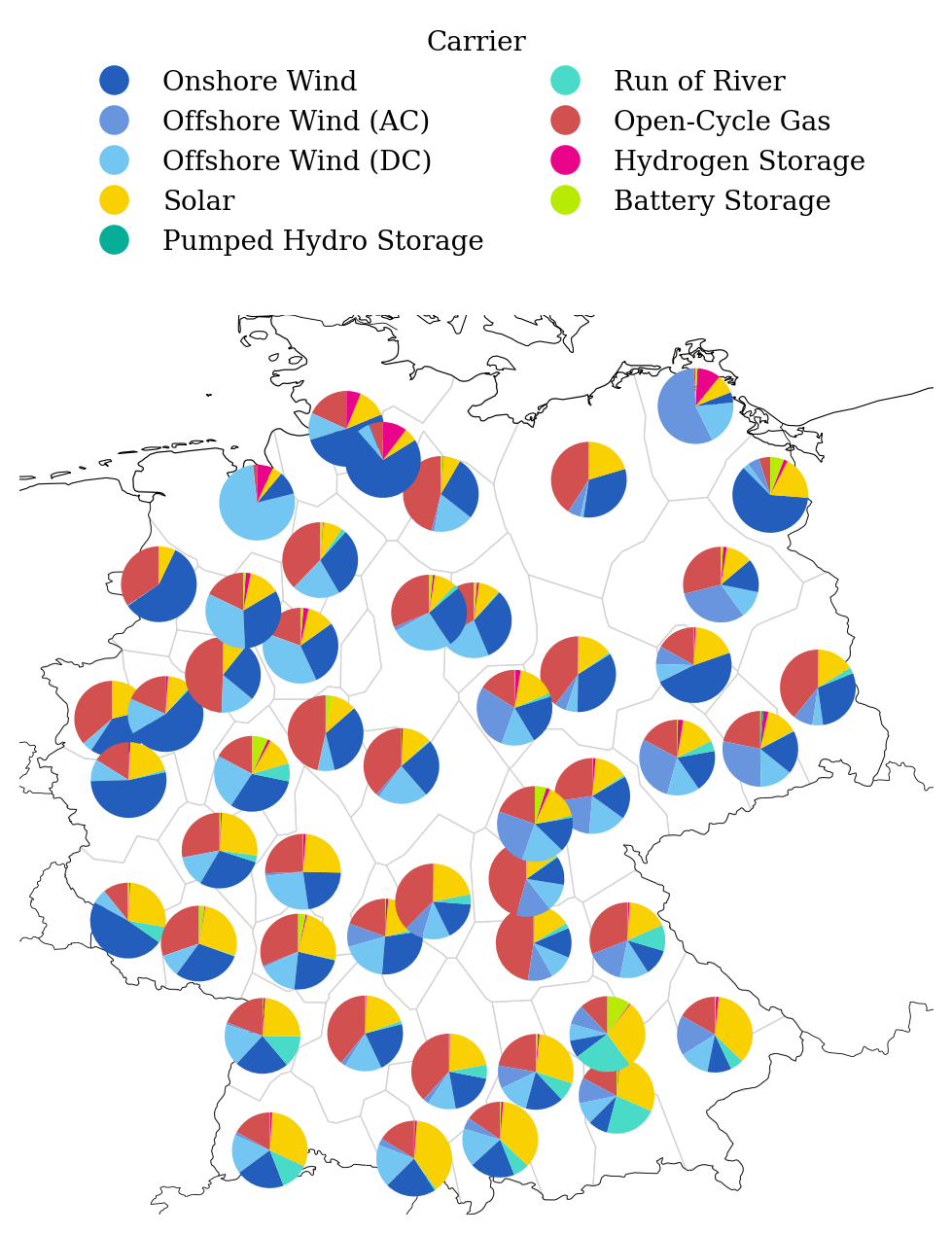}
    \caption{Average power mix per region calculated by Average Participation. Coastal regions are mainly supplied by local offshore and onshore wind farms. Their strong power injections additionally penetrate the network up to the southern border. In the middle and South, the supply is dominated by a combination of OCGT and solar power.}
    \label{fig:power_mix}
\end{figure}

\begin{figure*}
    \centering
    \begin{subfigure}[c]{.49\linewidth}
        \includegraphics[width=\linewidth]{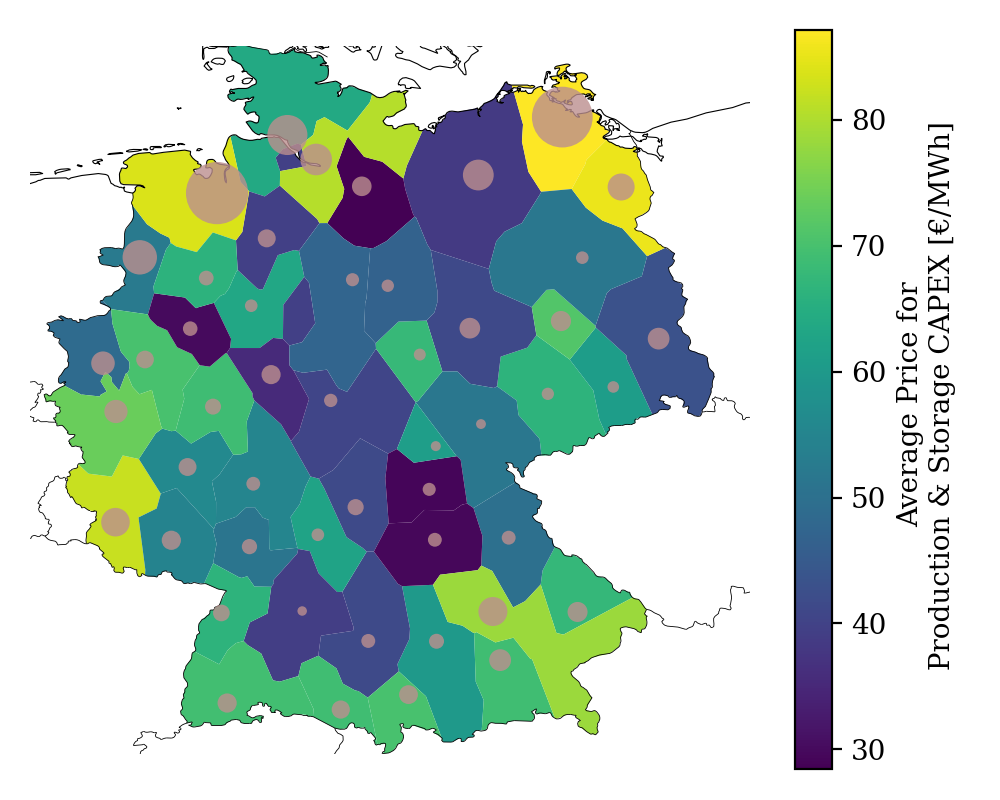}
        \subcaption{All production and storage technologies}
        \label{fig:total_capex}
    \end{subfigure}
    \begin{subfigure}[c]{.49\linewidth}
        \includegraphics[width=\linewidth]{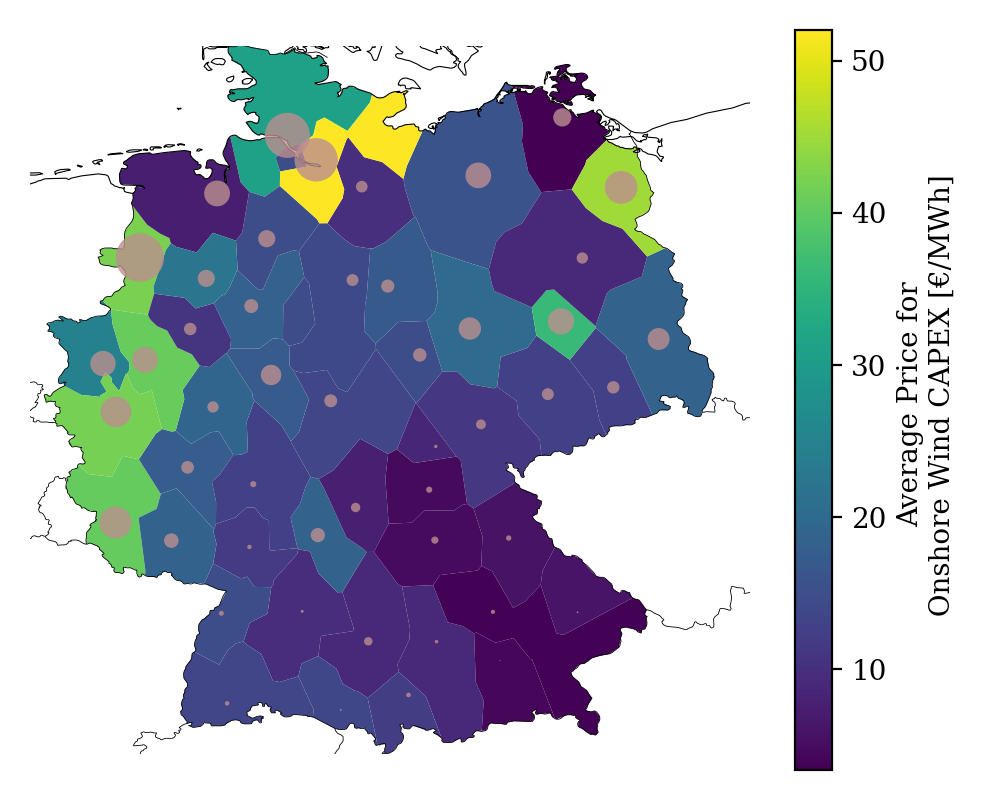}
        \subcaption{Onshore Wind}
        \label{fig:onshore_capex}
    \end{subfigure}
    \begin{subfigure}[c]{.49\linewidth}
        \includegraphics[width=\linewidth]{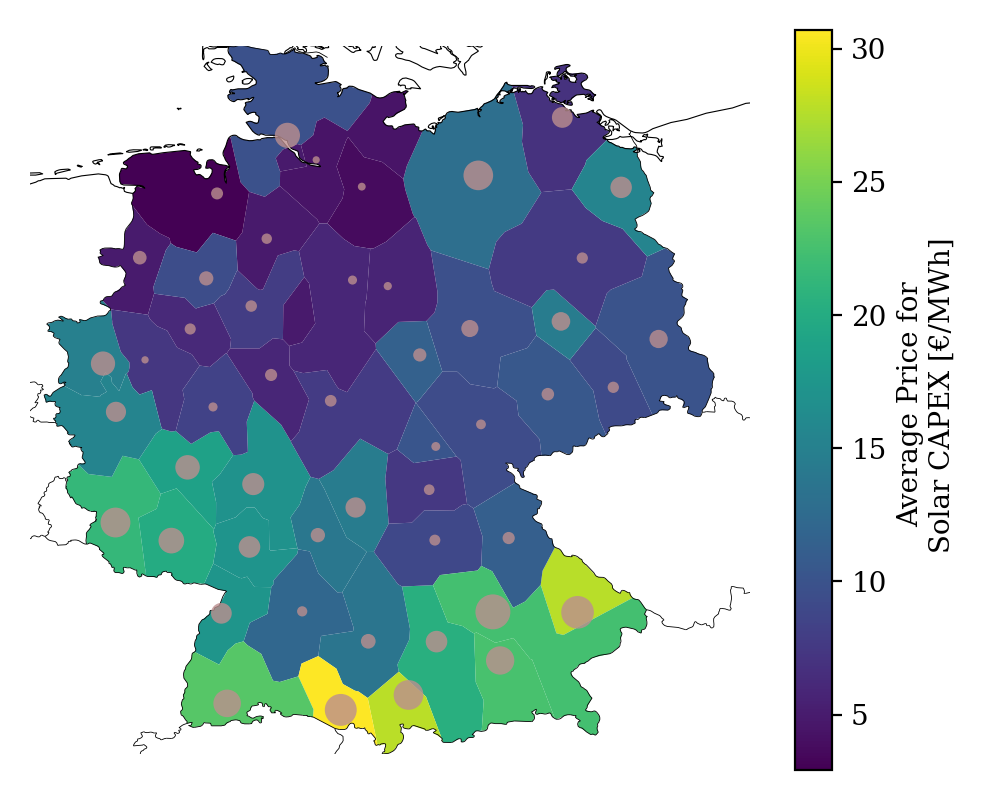}
        \subcaption{Solar}
        \label{fig:solar_capex}
    \end{subfigure}
    \begin{subfigure}[c]{.49\linewidth}
        \includegraphics[width=\linewidth]{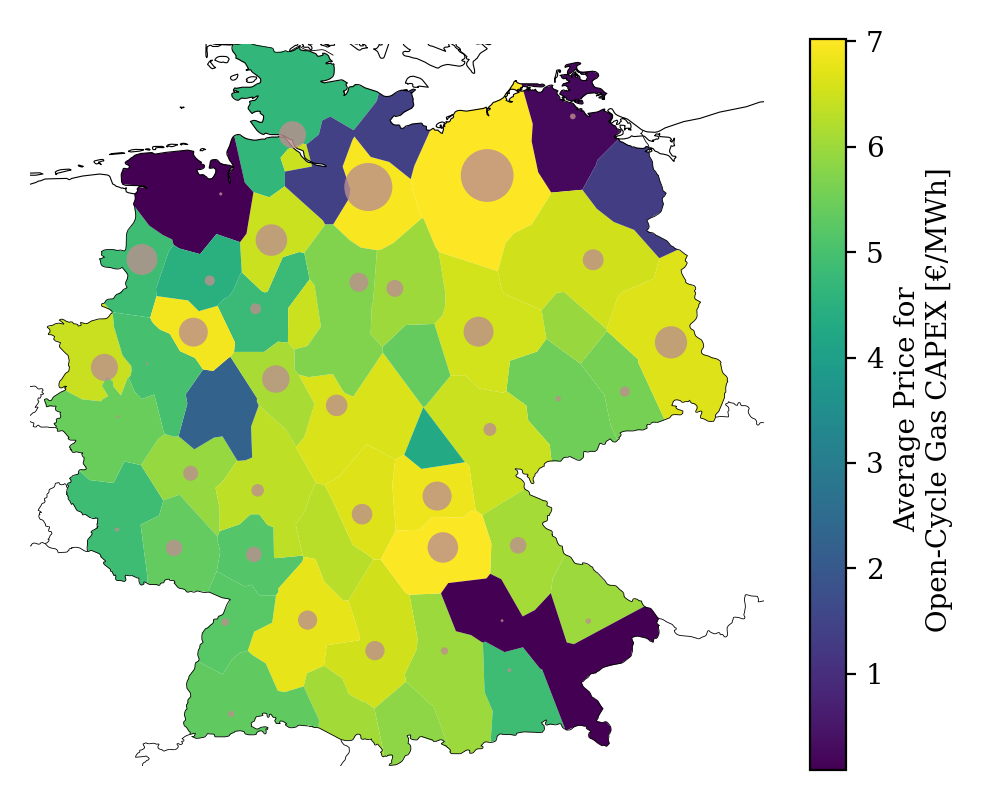}
        \subcaption{OCGT}
        \label{fig:ocgt_capex}
    \end{subfigure}
    \caption{Average \ac{CAPEX} allocation per MWh, $\sum_t \allocatecapex / \sum_t \demand$  for all production and storage assets (a), onshore wind (b), solar (c) and OCGT (d). Average allocated CAPEX per MWh within the regions are indicated by the color, the revenue per production asset is given by the size of the circles at the corresponding bus.}
    \label{fig:capex_price}
\end{figure*}

\begin{figure}
    \includegraphics[width=\linewidth]{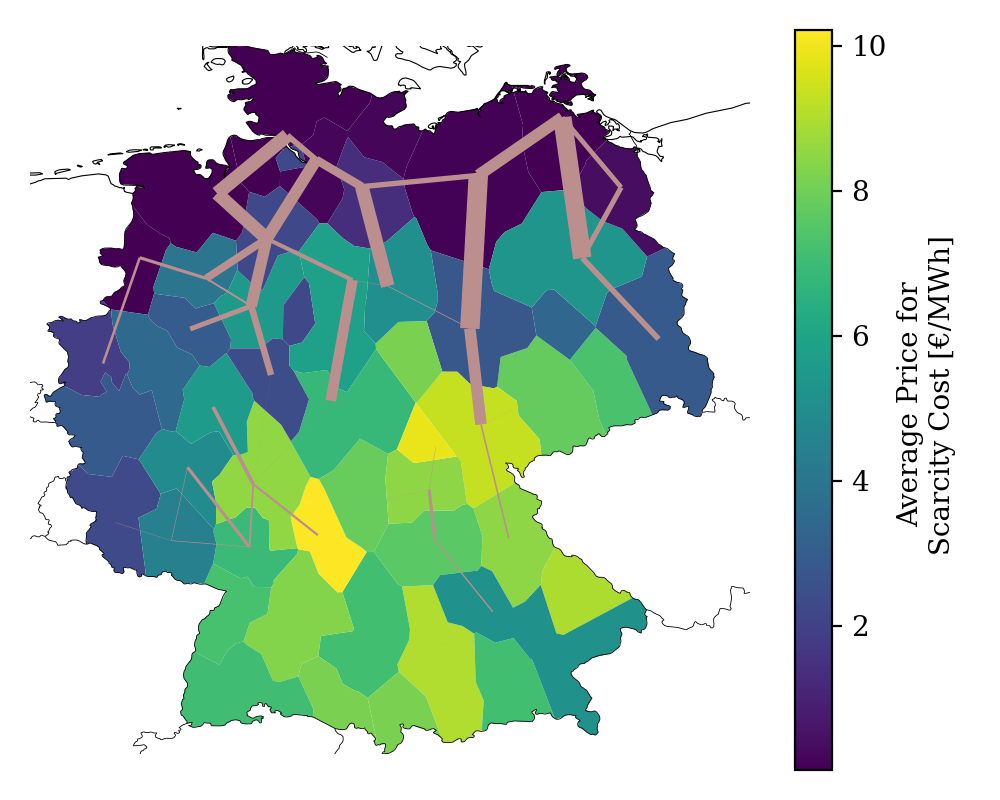}
    \caption{Average allocated transmission scarcity cost per consumed MWh, $\sum_t \scarcitycost_{n \rightarrow \ell, t} / \sum_t \demand $. This scarcity cost results from the upper transmission expansion limit of 25\%. The costs are indicated by the regional color.  The lines are drawn in proportion to revenue dedicated to scarcity cost. }
    \label{fig:branch_scarcity_price}
\end{figure}

\begin{figure}
    \vspace{2cm}
    \includegraphics[width=\linewidth]{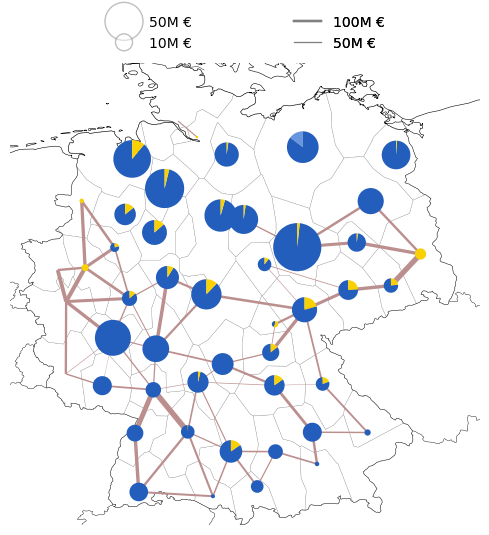}
    \caption{Total costs for subsidy $\subsidycost$ resulting from lower capacity expansion bounds (brownfield constraints). The figure shows the built infrastructure that does not gain back its CAPEX from its market revenue, but is only built due to lower capacity limits.}
    \label{fig:subsidy}
\end{figure}

\clearpage
\printbibliography

\end{document}

%% file: tables/de50/prices.tex
\begin{tabular}{lllr}
\toprule
     &    & o [\euro/MWh] &  c [k\,\euro/MW]$^*$ \\
{} & carrier &               &                      \\
\midrule
Generator & Open-Cycle Gas &       130.308 &               47.235 \\
     & Offshore Wind (AC) &         0.015 &              204.179 \\
     & Offshore Wind (DC) &         0.015 &              230.093 \\
     & Onshore Wind &         0.015 &              109.296 \\
     & Run of River &               &              270.941 \\
     & Solar &          0.01 &               55.064 \\
Storage & Hydrogen Storage &               &              224.739 \\
     & Pumped Hydro Storage &               &              160.627 \\
     & Battery Storage &               &              133.775 \\
Line & AC &               &                0.038 \\
     & DC &               &                0.070 \\
\bottomrule
\end{tabular}